\documentclass[11pt]{article}
\pdfoutput=1
\usepackage{amsmath,amssymb}
\usepackage{amsthm}
\usepackage{ascmac}
\usepackage{float}
\usepackage{bm}
\usepackage{graphicx}
\usepackage{comment}
\usepackage[all]{xy}
\usepackage{tikz-cd}
\usetikzlibrary{decorations.pathreplacing}
\usepackage[bookmarksnumbered=true,colorlinks=true,filecolor=blue,linkcolor=blue,urlcolor=blue,citecolor=red,linktocpage=true]{hyperref}

\allowdisplaybreaks

\numberwithin{equation}{section}

\def\cN{\mathcal{N}}

\def\cZ{\mathcal{Z}}

 \def\f {\frac}

\newcommand{\re}{\mathrm{e}}
\newcommand{\ri}{\mathrm{i}}
\newcommand{\rd}{\mathrm{d}}

\newcommand{\x}{\mathsf{x}}

\newcommand{\be}{\begin{equation}}
\newcommand{\ee}{\end{equation}}
\newcommand{\ba}{\begin{aligned}}
\newcommand{\ea}{\end{aligned}}

\usepackage{ulem}
\usepackage{dynkin-diagrams}
\newcommand{\rem}[1]{\textbf{\textcolor{blue}{(#1)}}}
\newcommand{\Tr}{\operatorname{Tr}}
\newcommand{\Str}{\operatorname{Str}}

\begin{document}

\renewcommand{\thefootnote}{\fnsymbol{footnote}}
\setcounter{page}{0}
\thispagestyle{empty}
\begin{flushright} \texttt{USTC-ICTS/PCFT-20-02} \end{flushright} 

\vskip2cm
\begin{center}

 {\LARGE \textbf{Topological Vertex/anti-Vertex\\ and\\[.5em] Supergroup Gauge Theory}} 
 \vskip1.5cm
{\large 
 {
 \sc Taro Kimura%
 \footnote{\href{mailto: taro.kimura@u-bourgogne.fr}{\tt taro.kimura@u-bourgogne.fr}}%
and
 \sc Yuji Sugimoto%
 \footnote{\href{mailto:sugimoto@ustc.edu.cn}{\tt sugimoto@ustc.edu.cn}}}

\vskip1.5cm
 \it $^{*}$Institut de Math\'ematiques de Bourgogne, UMR 5584, CNRS,\\
 Universit\'e Bourgogne Franche-Comt\'e, 21078 Dijon, France

\vskip.5cm
\it $^{\dagger}$NSFC-SFTP  Peng Huanwu Center for Fundamental Theory and Interdisciplinary Center for Theoretical  Study University of Science and Technology of China,\\  Hefei, Anhui 230026, China}
\end{center}

\vskip1cm
\begin{abstract} 
 We propose a new vertex formalism, called \textit{anti-refined topological vertex} (anti-vertex for short), to compute the generalized topological string amplitude, which gives rise to the supergroup gauge theory partition function.
 We show the one-to-many correspondence between the gauge theory and the Calabi--Yau geometry, which is peculiar to the supergroup theory, and the relation between the ordinary vertex formalism and the vertex/anti-vertex formalism through the analytic continuation.
\end{abstract}

\renewcommand{\thefootnote}{\arabic{footnote}}
\setcounter{footnote}{0}

\vfill\eject

\tableofcontents

\vspace{2em}

\hrule


\section{Introduction}\label{Intro}

Supersymmetric gauge theory plays a key role to understand a lot of aspects of string/M-theory. For example, the 4d/5d/6d supersymmetric gauge theories can be realized as Calabi--Yau compactification of the string/M-theory, so that we can study the string/M-theory duality by the supersymmetric gauge theories.
Especially, for the 5d $\cN=1$ SU($N$) supersymmetric gauge theories which can be also embedded into the string/M-theory, we can solve the low energy dynamics according to the Seiberg--Witten theory~\cite{Witten:1997sc}.

Certain classes of 5d $\cN=1$ SU($N$) supersymmetric gauge theories have realization using the $(p,q)$ 5-brane web diagram in type IIB superstring theory. Once we find the brane construction of the theory, its extension is possible by introducing additional stringy objects. For example, incorporating O$5$-planes, O$7$-planes, and $(p,q)$ 7-branes to the web diagram, one can construct the supersymmetric gauge theories with various types of the gauge group, such as Sp($N$), SO($N$), and exceptional groups.
The partition function of these theories can be calculated by the localization method in particular for the classical gauge groups~\cite{Nekrasov:2002qd,Marino:2004cn,Nekrasov:2004vw}. However, as a more efficient way, we can use the topological vertex techniques developed to calculate the partition function of the topological string theory on non-compact toric Calabi--Yau manifold~\cite{IKV, AKMV2005, AK}. Originally, the equivalence of the partition function of 5d $\cN=1$ SU($N$) supersymmetric gauge theory compactified on a circle $S^1$ and the toplogical string theory on non-compact toric Calabi--Yau manifold has been proposed in~\cite{KKV, KMV, DV, HIV} which is now known as {\it geometric engineering}.
After several efforts, we can calculate the partition function of the $5d$ $\cN=1$ supersymmetric gauge theories with several gauge groups we mentioned above. 
Moreover, quite interestingly, we can even express the 6d theory compactified on a torus by the $(p,q)$ 5-brane web with or without orientifold-plane in a sense that we can obtain the elliptic genus of the instanton moduli space engineered in M-theory from the web diagram by the topological vertex techniques~\cite{HIKLV, HKLGV, KTY, HKLY}.
Therefore, the web diagram of supersymmetric gauge theories and the topological vertex computation are compatible, and we can discuss several dualities pictorially and quantatively. For example, through the Hanany--Witten effect by moving $(p,q)$ 7-brane, we find a duality which can be checked by calculating the partition functions from the topological vertex.

Another interesting extension in the gauge theory is the supergroup gauge theories. In the type IIA construction, this gauge theory can be realized by adding {\it negative (ghost) branes}~\cite{OT, V2001, DHJV}.
For example, the U$(m|n)$ gauge theory can be realized by the $m$ D4-branes and the $n$ negative D4-branes suspended between two paralell NS5-branes (see also Fig.~\eqref{U(m|n)}).
Although the partition function of the supergroup gauge theory has been provided in~\cite{KP1} based on ADHM construction, neither the web diagram description nor topological vertex realization are given. 

In this paper, we propose the web diagram description of supergroup gauge theories by introducing {\it anti-trivalent graph}, and define {\it anti-topological vertex} as an extension of the topological vertex formalism giving the instanton partition functions of supergroup gauge theories. After giving the prescription how to use the new vertex, as a consistency check, we calculate the building blocks of the web diagram, and reproduce the instanton contributions for the vector multiplets, bifundamental multiplets, and fundamental multiplets given in~\cite{KP1}. 

Then, as an application, we check the expected \textit{ambiguity} of the web diagram description of supergroup gauge theories. For example, when we consider U$(2|1)$ gauge theory, there are more than two web diagram descriptions. We show that the partition functions calculated from these web diagrams are equivalent under suitable parameter correspondence. 
As another application, we consider the relation of the gauge theory with the supergroup and ordinary group which is also discussed in~\cite{DHJV,KP1}. For simplicity, in this application we consider the unrefined (anti-)topological vertex. Then, we find the explicit relation between anti-topological vertex and usual topological vertex. We also find the relation between the building block of the supergroup gauge theory and the ordinary gauge theory.

The organization of this paper is as follows. In Section~\ref{SGT}, we briefly review the supergroup gauge theories and their partition functions. In Section~\ref{formalism}, we propose the anti-refined topological vertex and corresponding anti-trivalent graph in the same way as the ordinary topological vertex and its trivalent graph description. Then, in Section~\ref{BSuCo}, we calculate the building blocks of the web diagram, and discuss the ambiguity of the web diagram description of U$(2|1)$ supergroup gauge theories. After that, in Section~\ref{sec:chain_gmtry}, we consider the web diagram describing the pure supergroup gauge theories, and discuss the relation between the gauge theory with the supergroup and usual group in the unrefined case. Finally we summarize our results and discuss some future works in Section~\ref{Disc}. In Appendix~\ref{ND}, we summarize some notations and useful formulae.


\section{Supergroup Gauge Theory}
\label{SGT}

In this Section, we briefly review the supergroup gauge theory and the associated supersymmetric localization to derive the partition function.

\subsection{Lagrangian and instantons}

Let $G$ be a Lie supergroup, e.g., $G = \mathrm{U}(n_+|n_-)$.
The dynamical degrees of freedom of the supergroup gauge theory is the gauge field, which transforms under the supergroup action, $A_\mu \to g A_\mu g^{-1} + g \partial_\mu g^{-1}$ with $g \in G$.
Then the $G$-gauge invariant Yang--Mills action is given by
\begin{align}
 S_\text{YM} = \frac{1}{2 g^2_\text{YM}} \int \rd^4 x \, \Str F_{\mu\nu} F^{\mu\nu}
\end{align}
where ``Str'' is the supertrace operation, which provides the Killing form for the corresponding Lie superalgebra:
For example, the supertrace over the graded vector space $\mathbb{C}^{n_+|n_-} = \mathbb{C}^{n_+} \oplus \mathbb{C}^{n_-}$ is given by $\Str_{\mathbb{C}^{n_+|n_-}} = \Tr_{\mathbb{C}^{n_+}} - \Tr_{\mathbb{C}^{n_-}}$.
In particular, for $G = \mathrm{U}(n_+|n_-)$, the Yang--Mills action consists of positive and negative parts,
\begin{align}
 S_\text{YM} =
 \frac{1}{2 g^2_\text{YM}} \int \rd^4 x \, \Tr_{\mathbb{C}^{n_+}} \left(F_{\mu\nu} F^{\mu\nu}\right)^+ - \frac{1}{2 g^2_\text{YM}} \int \rd^4 x \, \Tr_{\mathbb{C}^{n_-}} \left(F_{\mu\nu} F^{\mu\nu}\right)^-
\end{align}
where $\left(F_{\mu\nu} F^{\mu\nu}\right)^\pm$ schematically denotes the positive/negative contribution of the Lagrangian.
We immediately notice that this action is unbounded since the kinetic term of the negative part has a wrong sign.
In other words, the $\mathrm{U}(n_+|n_-)$ gauge theory is interpreted as a quiver gauge theory with two gauge nodes, $(G_+, G_-) = (\mathrm{U}(n_+), \mathrm{U}(n_-))$, where the off-diagonal components of the gauge field play a role of the bifundamental matters~\cite{DHJV,Gaiotto:2008sd,Drukker:2009hy,MP,Mikhaylov:2014aoa}.
The supergroup gauge invariance imposes a constraint on their coupling constatnts $(\tau_+,\tau_-) = (\tau, - \tau)$, where $\tau$ is the complexified coupling constant $\displaystyle \tau = \frac{\theta}{2\pi} + \frac{4 \pi \ri}{g^2_\text{YM}}$ associated with the $G$-invariant $\theta$-term,
\begin{align}
 S_\theta = \frac{\theta}{16 \pi^2} \int \rd^4 x \, \Str F_{\mu\nu} \tilde{F}^{\mu\nu}
 \, .
\end{align}
The imaginary part of the physical coupling should be positive, $\operatorname{Im}\tau = 4 \pi / g^2_\text{YM} > 0$.
Once assuming $\operatorname{Im} \tau_+ > 0$, the other must be $\operatorname{Im}\tau_- < 0$, which is unphysical, due to the supergroup condition.

As well as the ordinary gauge theory, the $\theta$-term and the associated topological configuration, namely the instanton, play an important role in the non-perturbative aspects of the supergroup gauge theory.
The ADHM analysis for the supergroup gauge theory shows that we need two non-negative integers to parametrize the topological charge of the instanton~\cite{KP1},
\begin{align}
 - \frac{1}{16 \pi^2} \int \rd^4 x \, \Str F_{\mu\nu} \tilde{F}^{\mu\nu} = k_+ - k_-
 \, ,
\end{align}
where $k_\pm$ is interpreted as the positive/negative instanton number.

\subsection{Supersymmetric localization}\label{sec:loc}

In the presence of the spacetime supersymmetry, we can further study the non-perturbative aspects of gauge theory.
In particular, 4d $\cN = 2$ (5d $\cN = 1$) gauge theory partition function is obtained through the instanton counting with the $\Omega$-background~\cite{Nekrasov:2002qd} together with the supersymmetric localization.
In the case of the supergroup gauge theory, it turns out to be a double sum over the positive and negative instantons~\cite{KP1},
\begin{align}
 \cZ = \sum_{k_\pm = 0}^\infty \mathfrak{q}^{k_+ - k_-} \, \cZ_{k_+|k_-}
\end{align}
where the instanton fugacity $\mathfrak{q}$ is given by the complexified coupling $\mathfrak{q} = \exp \left( 2 \pi \ri \tau \right)$.
The $(k_+|k_-)$-instanton contribution for $\mathrm{U}(n_+|n_-)$ gauge theory is given by 
\begin{align}
 \cZ_{k_+|k_-} = \sum_{|\vec{\lambda}^\pm| = k_\pm} \cZ_{\vec{\lambda}^\pm}
 \, .
\end{align}
Here we define two sets of partitions
\begin{align}
 \vec{\lambda}^+ = (\lambda^+_1, \ldots, \lambda_{n_+}^+)
 \, , \qquad
 \vec{\lambda}^- = (\lambda^-_1, \ldots, \lambda_{n_-}^-)
 \, ,
\end{align}
where each $\lambda^\pm_\alpha$ is a partition obeying the non-increasing condition
\begin{align}
 \lambda^\pm_\alpha = (\lambda_{\alpha,1}^\pm \ge \lambda_{\alpha,2}^\pm \ge \cdots \ge 0) \in \mathbb{Z}_{\ge 0}^\infty
\end{align}
and
\begin{align}
 |\vec{\lambda}^+| = \sum_{\alpha=1}^{n_+} \sum_{i=1}^\infty \lambda_{\alpha,i}^+
 \, , \qquad
 |\vec{\lambda}^-| = \sum_{\alpha=1}^{n_-} \sum_{i=1}^\infty \lambda_{\alpha,i}^-
 \, .
\end{align}
The contribution $\cZ_{\vec{\lambda}^\pm}$ consists of the vector, fundamental and antifundamental hyper multiplet factors, evaluated with the super instanton configuration $(\vec{\lambda}^+,\vec{\lambda}^-)$,%
\footnote{%
One can also impose the Chern--Simons term in particular for 5d gauge theory, which we do not consider in this paper for simplicity.
}
\begin{align}
 \cZ_{\vec{\lambda}^\pm} = Z^\text{vect} Z^\text{f} Z^\text{af}
 \, .
\end{align}
For quiver gauge theory, involving plural gauge nodes, there also exists the bifundamental matter contribution, connecting different gauge nodes.
The explicit forms of these contributions are shown in the following.

\paragraph{Vector multiplet}

We first consider the vector multiplet contribution.
In order to write down the formula, we define a set of the dynamical $x$-variables,%
\footnote{%
In the following, the $\Omega$-background parameter $(\epsilon_1,\epsilon_2)$ will be identified with the refined string coupling constatns, $(q,t) = (\re^{\epsilon_1}, \re^{-\epsilon_2})$. We also use the notation $\epsilon_+ := \epsilon_1 + \epsilon_2 = \log (q/t)$, which becomes zero in the unrefined situation.
}
which characterizes the instanton configuration instead of the partition,
\begin{align}
 \mathcal{X}^\pm & = \{ x_{\alpha,i}^\pm = \re^{a_\alpha^\pm \pm \epsilon_2 \lambda_{\alpha,i}^\pm \pm \epsilon_1 (i - 1)} \}_{\alpha = 1,\ldots,n_\pm, i = 1,\ldots,\infty}
 \, .
\end{align}
This implies that the $\Omega$-background parameters are flipped for the negative node similarly to the coupling constant~\cite{KP1}. 
Then the Chern supercharacter of the corresponding universal sheaf on the instanton moduli space is written in terms of the $x$-variables,
\begin{align}
 \operatorname{sch} \mathbf{Y} = \operatorname{ch} \mathbf{Y}^+ - \operatorname{ch} \mathbf{Y}^-
\end{align}
where
\begin{align}
 \operatorname{ch} \mathbf{Y}^\pm = (1 - \re^{\pm \epsilon_1}) \sum_{x \in \mathcal{X}^\pm} x .
\end{align}

The supercharacter of the virtual class for the vector multiplet is defined
\begin{align}
 \operatorname{sch} \mathbf{V}
 = \frac{\operatorname{sch} \mathbf{Y}^\vee \operatorname{sch} \mathbf{Y}}{(1 - \re^{\epsilon_1})(1 - \re^{\epsilon_2})}
 = \sum_{\sigma,\sigma' = \pm} \sigma\sigma' \operatorname{ch} \mathbf{V}_{\sigma\sigma'}
\end{align}
with the dual denoted by $\mathbf{Y}^\vee$, and each contribution has an explicit form in terms of the $x$-variables
\begin{subequations}
 \begin{align}
  & \operatorname{ch} \mathbf{V}_{++} = \frac{1 - \re^{-\epsilon_1}}{1 - \re^{\epsilon_2}} \sum_{\substack{(x,x') \in \mathcal{X}^+ \times \mathcal{X}^+\\ x \neq x'}} \frac{x'}{x}
  \, , \quad
  & \operatorname{ch} \mathbf{V}_{+-} = \re^{-\epsilon_+} \frac{1 - \re^{-\epsilon_1}}{1 - \re^{-\epsilon_2}} \sum_{(x,x') \in \mathcal{X}^+ \times \mathcal{X}^-} \frac{x'}{x}
  \, , \\
  & \operatorname{ch} \mathbf{V}_{-+} = \frac{1 - \re^{\epsilon_1}}{1 - \re^{\epsilon_2}} \sum_{(x,x') \in \mathcal{X}^- \times \mathcal{X}^+} \frac{x'}{x}
  \, , \quad
  & \operatorname{ch} \mathbf{V}_{--} = \frac{1 - \re^{-\epsilon_1}}{1 - \re^{\epsilon_2}} \sum_{\substack{(x,x') \in \mathcal{X}^- \times \mathcal{X}^-\\ x \neq x'}} \frac{x'}{x}
  \, .  
 \end{align}
\end{subequations}
Then the full partition function contribution for the vector multiplet is given by
\begin{align}
 Z^\text{vect} = \mathbb{I}[\mathbf{V}] = \prod_{\sigma,\sigma'=\pm} Z^\text{vect}_{\sigma\sigma'}
\end{align}
where we apply the index defined, for a given character $\operatorname{ch} \mathbf{X} = \sum_{X} x$,%
\footnote{%
The 6d $\mathcal{N} = (1,0)$ gauge theory partition function compactified on a torus with the modulus $\tau$ is similarly obtained with the elliptic analog of the index,
\begin{align}
 \mathbb{I}_p[\mathbf{X}] = \prod_{x \in X} \theta(x^{-1};p)
\end{align}
with the elliptic nome $p = \exp \left( 2 \pi \ri \tau \right)$ and the theta function defined
\begin{align}
 \theta(z;p) = (z;p)_\infty (p z^{-1};p)_\infty
 \, .
\end{align}
See, for example, \cite{Kimura:2016dys} for details.
}
\begin{align}
 \mathbb{I}[\mathbf{X}] = \prod_{x \in X} \left( 1 - x^{-1}\right)
 \, .
\end{align}
Each contribution in terms of the $x$-variables has an explicit form of
\begin{subequations}
 \begin{align}
  & Z_{++}^\text{vect} = \prod_{\substack{(x,x') \in \mathcal{X}^+ \times \mathcal{X}^+ \\ x \neq x'}} \frac{(\re^{\epsilon_+} x/x';\re^{\epsilon_2})_\infty}{(\re^{\epsilon_2} x/x';\re^{\epsilon_2})_\infty}
  & Z_{+-}^\text{vect} = \prod_{(x,x') \in \mathcal{X}^+ \times \mathcal{X}^-} \frac{(\re^{\epsilon_+ + \epsilon_1} x/x';\re^{\epsilon_2})_\infty}{(\re^{\epsilon_+} x/x';\re^{\epsilon_2})_\infty}
  \\
  & Z_{-+}^\text{vect} = \prod_{(x,x') \in \mathcal{X}^- \times \mathcal{X}^+} \frac{(\re^{\epsilon_2} x/x';\re^{\epsilon_2})_\infty}{(\re^{-\epsilon_1+\epsilon_2} x/x';\re^{\epsilon_2})_\infty}
  & Z_{--}^\text{vect} = \prod_{\substack{(x,x') \in \mathcal{X}^- \times \mathcal{X}^- \\ x \neq x'}} \frac{(\re^{\epsilon_+} x/x';\re^{\epsilon_2})_\infty}{(\re^{\epsilon_2} x/x';\re^{\epsilon_2})_\infty}  
 \end{align}
\end{subequations}
with the $q$-shifted factorial ($q$-Pochhammer) symbol
\begin{align}
 (z;q)_n = \prod_{k=0}^{n-1} (1 - z q^k)
 \, .
\end{align}

\paragraph{Bifundamental hypermultiplet}

The bifundamental hypermultiplet contribution is similarly formulated for quiver gauge theory.
Let $\Gamma$ be a quiver $\Gamma = (\Gamma_0, \Gamma_1)$ with sets of the gauge nodes $\Gamma_0$ and the edges $\Gamma_1$.
We define the Chern supercharacter for the virtual class for the bifundamental hypermultiplet,
\begin{align}
 \operatorname{sch} \mathbf{H}_{e} = - \re^{m_{e}} \frac{\operatorname{sch} \mathbf{Y}^\vee_{s(e)} \operatorname{sch} \mathbf{Y}_{t(e)}}{(1 - \re^{\epsilon_1})(1 - \re^{\epsilon_2})}
\end{align}
where $e \in \Gamma_1$ is the edge connecting the source node $s(e)$ and the target node $t(e)$;
$m_e$ is the corresponding bifundamental mass parameter;
$(\mathbf{Y}_k)_{k \in \Gamma_0}$ is a collection of the universal sheaves assigned to each node $k \in \Gamma_0$.
Define sets of $x$-variables $\mathcal{X}_k$ for the node $k \in \Gamma_0$.
Then the full partition function contribution is given by
\begin{align}
 Z^\text{bifund}_e = \mathbb{I}[\mathbf{H}_e] = \prod_{\sigma,\sigma'=\pm} Z^\text{bifund}_{e,\sigma\sigma'}
\end{align}
where
\begin{subequations}
 \begin{align}
  & Z^\text{bifund}_{e,++} = \prod_{\substack{(x,x') \in \mathcal{X}_{s(e)}^+ \times \mathcal{X}_{t(e)}^+ \\ x \neq x'}} \frac{(\re^{-m_e + \epsilon_2} x/x';\re^{\epsilon_2} )_\infty}{(\re^{-m_e + \epsilon_+} x/x';\re^{\epsilon_2} )_\infty}
  \, ,
  \\
  & Z^\text{bifund}_{e,+-} = \prod_{(x,x') \in \mathcal{X}_{s(e)}^+ \times \mathcal{X}_{t(e)}^-} \frac{(\re^{-m_e + \epsilon_+} x/x';\re^{\epsilon_2} )_\infty}{(\re^{-m_e + \epsilon_+ + \epsilon_1} x/x';\re^{\epsilon_2} )_\infty}
  \, ,
  \\
  & Z^\text{bifund}_{e,-+} = \prod_{(x,x') \in \mathcal{X}_{s(e)}^- \times \mathcal{X}_{t(e)}^+} \frac{(\re^{-m_e - \epsilon_1 + \epsilon_2} x/x';\re^{\epsilon_2} )_\infty}{(\re^{-m_e + \epsilon_2} x/x';\re^{\epsilon_2} )_\infty}
  \, ,
  \\
  & Z^\text{bifund}_{e,--} = \prod_{\substack{(x,x') \in \mathcal{X}_{s(e)}^- \times \mathcal{X}_{t(e)}^- \\ x \neq x'}} \frac{(\re^{-m_e + \epsilon_2} x/x';\re^{\epsilon_2} )_\infty}{(\re^{-m_e + \epsilon_+} x/x';\re^{\epsilon_2} )_\infty}
  \, .
 \end{align}
\end{subequations}

\paragraph{Fundamental hypermultiplet}

For the fundamental and antifundamental hypermultiplets, we define
\begin{subequations}
\begin{align}
 \operatorname{sch} \mathbf{H}^\text{f} & = - \frac{\operatorname{sch} \mathbf{Y}^\vee \operatorname{sch} \mathbf{M}}{(1 - \re^{\epsilon_1})(1 - \re^{\epsilon_2})}
 \\
 \operatorname{sch} \mathbf{H}^\text{af} & = - \frac{\operatorname{sch} \widetilde{\mathbf{M}}^\vee \operatorname{sch} \mathbf{Y}^\vee}{(1 - \re^{\epsilon_1})(1 - \re^{\epsilon_2})} 
\end{align}
\end{subequations}
where the Chern roots for the matter sheaves are given by the fundamental and the antifundamental mass parameters, namely the equivariant parameters associated with the flavor symmetry,
\begin{subequations}
\begin{align}
 \operatorname{sch} \mathbf{M} & = \sum_{f=1}^{n^\text{f}_+} \re^{m_f^+} - \sum_{f=1}^{n^\text{f}_-} \re^{m_f^-}
 \\
 \operatorname{sch} \widetilde{\mathbf{M}} & = \sum_{f=1}^{n^\text{af}_+} \re^{\tilde{m}_f^+} - \sum_{f=1}^{n^\text{af}_-} \re^{\tilde{m}_f^-} 
\end{align}
\end{subequations}
Then the full partition function contribution is given by the index
\begin{align}
 Z^\text{(a)f} = \mathbb{I}[\mathbf{H}^\text{(a)f}]
 \, .
\end{align}
We remark that these are also obtained from the bifundamental contribution by freezing either the source or the target gauge node.
See \cite{KP1} for the explicit forms of the full partition function contribution in this case.

\section{Anti-Refined Topological Vertex}
\label{formalism}

\subsection{The vertex}
Before introducing the new vertex, we briefly review the ordinary refined topological vertex $C_{\lambda \mu \nu}(t,q)$ defined in~\cite{IKV, AKMV2005, AK},
\be
\ba
C_{\lambda \mu \nu}(t,q) = t^{-\f{1}{2}||\mu^{\rm T}||^2}q^{\f{1}{2}(||\mu||^2 + ||\nu||^2)} \tilde{Z}_\nu (t, q) \sum_{\eta} \left( \f{q}{t} \right)^{\f{1}{2}(|\eta|+|\lambda|-|\mu|)} s_{\lambda^{\rm T}/\eta}(t^{-\rho} q^{-\nu}) s_{\mu/\eta}(q^{-\rho} t^{-\nu^{\rm T}}),
\ea
\ee
where $s_{\mu/\eta}(x)$ is the {skew-Schur function}, and $\lambda,\mu$ and $\nu$ denote the partitions (Young diagrams), characterizing the boundary condition of the vertex.
We summarize the notations, definitions, and useful formulae in Appendix~\ref{ND}. 
The refined topological vertex is depicted as a trivalent vertex ordered in a clockwise way:
\begin{align}
 \begin{tikzpicture}[thick,baseline=(current bounding box.center)]
\filldraw[fill=white,draw=black] node[above left = 10mm]{$C_{\lambda \mu \nu}(t,q)~~:=$};
\draw[-latex] (1,1) -- (0,0) node[below left = -1mm] {$\lambda$} node[right = 2mm] {$t$}  node[above left = -1mm] {$(-1,-1)$} ;
\draw[-latex] (1,1) -- (1,2.4) node[above=0.5mm] {$\mu$}  node[below right = 0mm] {$q$} node[below left = 0mm] {$(0,1)$} ;
\draw[-latex] (1,1) -- (2.4,1) node[below left=0.4mm] {$\nu$} node[above right = -0.5mm] {$(1,0)$} ;
 \end{tikzpicture}
\end{align}
where the partitions $\lambda,\mu$ and $\nu$, are assigned to the legs, and two of them have the argument, $t$ or $q$. We also assign the charge vectors for them. The leg without argument is called the \textit{preferred direction}. By gluing vertices, we can calculate the refined topological string amplitude on the non-compact toric, and some of non-toric Calabi--Yau geometries. 
The gluing rule for two vertices is to sum up all the possible partitions (namely, the boundary conditions) with the weight $(-Q)^{|\mu|}f_\mu^{\mathfrak{n}}$ or $(-Q)^{|\mu|} \tilde{f}_\mu^{\mathfrak{n}}$,
\begin{subequations}\label{uglue}
\begin{align}
 &\sum_{\mu}C_{\lambda_1 \nu_1 \mu}(t,q) C_{\lambda_2  \nu_2 \mu^{\rm T}}(q,t) (-Q)^{|\mu|} \tilde{f}_{\mu}^{\mathfrak{n}}(q,t) \quad \text{for preferred direction,}
\\
&\sum_{\mu}C_{\lambda_1 \mu \nu_1}(t,q) C_{\lambda_2 \mu^{\rm T} \nu_2}(q,t) (-Q)^{|\mu|}f_{\mu}^{\mathfrak{n}}(q,t) \quad \text{for other directions,}
\end{align}
\end{subequations}
where $f_\mu(t,q)$ and $\tilde{f}_\mu(t,q)$ are the framing factors,
\be
\ba
f_\mu (t,q) = (-1)^{|\mu|} t^{\f{||\mu^{\rm T}||^2}{2}} q^{-\f{||\mu||^2}{2}},
&&
\tilde{f}_\mu (t,q) = (-1)^{|\mu|} \left(\f{t}{q}\right)^{\f{|\mu|}{2}} t^{\f{||\mu^{\rm T}||^2}{2}} q^{-\f{||\mu||^2}{2}},
\ea
\ee
whose choice depends on which two legs are used to glue together: We use $f_\mu(t,q)$ to glue along the preferred directions, otherwise we use $\tilde{f}_\mu(t,q)$. The weight $Q$ is the K\"ahler parameter assigned to the edge.  The integer $\mathfrak{n}$ is given by the wedge product of two charge vectors, $\mathfrak{n}=v_2 \wedge v_1$, as follows:
\begin{align}
\begin{tikzpicture}[thick,baseline=(current bounding box.center)]
\footnotesize
\draw[-latex] (1,1) -- (0.3,0) node[ right= 1mm] {$\nu_1,t$} node[right = -20mm] {$v_1=(p_1,q_1)$} ;
\draw[-latex] (1,1) -- (-0.2,2) node[below=2mm] {$\lambda_1,q$}  node[below right = 0mm] {} ;
\draw[-latex] (1,1) -- (1.4,1) ;  \draw[dotted] (1.4,1) -- (2.0,1) ;  \draw[-latex] (2,1) -- (2.4,1) ;
\coordinate (O) at (1,0.7) node at (O) [right=2mm] {$Q,\tilde{f}^{\mathfrak{n}}_\mu$}; 
\coordinate (O) at (1,1.3) node at (O) [right=4.3mm] {$\mu$}; 
\draw[-latex] (2.4,1) -- (3,2) node[left = 1mm] {$\nu_2,q$} node[below right = 0mm] {$v_2=(p_2,q_2)$} ;
\draw[-latex] (2.4,1) -- (2.6,-0.2) node[above right = -0.5mm] {$\lambda_2,t$} node[right = 2mm] {} ;
\draw[-latex] (8,1) -- (7.3,0) node[ right= 1mm] {$\nu_1$} node[right = -20mm] {$v_1=(p_1,q_1)$} ;
\draw[-latex] (8,1) -- (6.8,2) node[below=2mm] {$\lambda_1,t$}  node[below right = 0mm] {} ;
\draw[-latex] (8,1) -- (8.4,1) ;  \draw[dotted] (8.4,1) -- (9.0,1) ;  \draw[-latex] (9,1) -- (9.4,1) ;
\coordinate (O) at (8,0.7) node at (O) [right=3mm] {$Q,f^{\mathfrak{n}}_\mu$}; 
\coordinate (O) at (8,1.2) node at (O) [right=0mm] {$q$}; 
\coordinate (O) at (8,1.3) node at (O) [right=4.3mm] {$\mu$}; 
\coordinate (O) at (9,1.2) node at (O) [right=0mm] {$t$}; 
\draw[-latex] (9.4,1) -- (10,2) node[left = 1mm] {$\nu_2$} node[below right = 0mm] {$v_2=(p_2,q_2)$} ;
\draw[-latex] (9.4,1) -- (9.6,-0.2) node[above right = -0.5mm] {$\lambda_2,q$} node[right = 2mm] {} ;
 \end{tikzpicture}
\end{align}

\subsection{The anti-vertex}

Now we define a new vertex, that we call \textit{anti-refined topological vertex} (anti-vertex for short), denoted by $\bar{C}_{\lambda \mu \nu}(t,q)$,
\be
\ba
\bar{C}_{\lambda \mu \nu}(t,q) = t^{-\f{1}{2}||\mu^{\rm T}||^2}q^{\f{1}{2}(||\mu||^2-||\nu||^2)} \tilde{Z}_\nu (t^{-1}, q^{-1}) \sum_{\eta} \left( \f{q}{t} \right)^{\f{1}{2}(|\eta|+|\lambda|-|\mu|)} s_{\lambda^{\rm T}/\eta}(t^\rho q^\nu) s_{\mu/\eta}(q^\rho t^{\nu^{\rm T}}).
\ea
\ee
This anti-vertex is essentially the same as the ordinary vertex, but with the replacement $(q,t) \leftrightarrow (q^{-1},t^{-1})$, which corresponds to the flip of the coupling constant $g_s \leftrightarrow - g_s$ where $q = \re^{-g_s}$.
This is the reason why we call this the anti-vertex since it has the negative coupling~\cite{V2001}.
This is consistent with the gauge theory perspective since these parameters are interpreted as the $\Omega$-background parameter $(q,t) = (\re^{\epsilon_1}, \re^{-\epsilon_2})$, and a similar flip, $(\epsilon_1,\epsilon_2) \leftrightarrow (-\epsilon_1,-\epsilon_2)$, is necessary for the supergroup gauge theory as seen in Sec.~~\ref{sec:loc}.

Similarly to the usual refined topological vertex, we introduce the graphical description for the anti-vertex as follows:
\begin{align}
 \begin{tikzpicture}[thick,baseline=(current bounding box.center)]
\filldraw[fill=white,draw=black] node[above left = 10mm]{$\bar{C}_{\lambda \mu \nu}(t,q)~~:=$};
\draw[-latex,dashed] (1,1) -- (0,0) node[below left = -1mm] {$\lambda$} node[right = 2mm] {$t$} ;
\draw[-latex,dashed] (1,1) -- (1,2.4) node[above=0.5mm] {$\mu$}  node[below right = 0mm] {$q$};
\draw[-latex,dashed] (1,1) -- (2.4,1) node[right] {$\nu$};
 \end{tikzpicture}
\end{align}
The gluing rule between the anti-vertex and the usual vertex, and two anti-vertices are the same as usual since the framing part of the anti-vertex except for the one along preferred direction, $t^{-\f{1}{2}||\mu^{\rm T}||^2}q^{\f{1}{2}||\mu||}$, and the prefactor in the summation, $\left( \f{q}{t} \right)^{\f{1}{2}(|\eta|+|\lambda|-|\mu|)} $, are the same as the ordinary refined topological vertex. From these rules, we will calculate the topological string amplitudes, which gives rise to the supergroup gauge theory partition functions.


\section{Geometric Engineering of Supergroup Gauge Theories with Matters}\label{BSuCo}

\subsection{The Building Block} 
To begin with, here we consider following chain geometry:
\begin{align}
\begin{tikzpicture}[thick,baseline=(current bounding box.center)]
\footnotesize
\coordinate (O) at (5.5,-2) node at (O) [left=5mm] {$\cZ_{\{ \nu_i \} , \{ \nu'_i \} }^{\text{build}} =~$}  ;
\draw[] (5.5,1) -- (5.5,1.5) ;
\draw[] (5,1) -- (5.5,1) node[left=4mm] {$\nu_1$};
\draw[] (6,0.5) -- (5.5,1) ;  \coordinate (O) at (5.6,0.7) node at (O) [right=-4mm] {$Q_1$}  ;
\draw[] (6.5,0.5) -- (6,0.5) node[right =5mm] {${\nu'}^{\rm{T}}_1$} ;
\draw[] (6,0.5) -- (6,0) ;  \coordinate (O) at (6,0.25) node at (O) [right=0mm] {$Q'_1$} ;
\draw[] (6,0) -- (5.5,0)  node[left=0mm] {$\nu_2$} ;
\draw[] (6,0) -- (6.5,-0.5) ; \coordinate (O) at (6.1,-0.3) node at (O) [right=-4mm] {$Q_2$}  ;
\draw[] (6.5,-0.5) -- (7,-0.5)  node[right=0mm] {${\nu'}^{\rm{T}}_2$} ;
\draw[] (6.5,-0.5) -- (6.5,-0.8) ; 
\draw[dotted] (6.5,-0.8)  -- (6.5,-1.2) ; 
\draw[] (6.5,-1.2) -- (6.5,-1.5) ; 
\draw[] (6.5,-1.5) -- (6,-1.5) node[left=-1mm] {$\nu_m$};
\draw[] (6.5,-1.5) -- (6.75,-1.75) ; 
  \coordinate (O) at (6.5,-1.8) node at (O) [right=-4mm] {$Q_m$}  ;
\draw[dashed]  (6.75,-1.75)  -- (7,-2) ;
\draw[dashed]  (7,-2)  -- (7.5,-2) node[right=0mm] {${\nu'}^{\rm{T}}_{m}$} ;
\draw[dashed]  (7,-2)  -- (7,-2.5) ;  \coordinate (O) at (7,-2.3) node at (O) [right=0mm] {$Q_m'$}  ;
\draw[dashed]  (7,-2.5)  -- (6.5,-2.5) node[left=-1mm] {$\nu_{m+1}$};
\draw[dashed]  (7,-2.5)  -- (7.5,-3) ;  \coordinate (O) at (6.7,-2.8) node at (O) [right=-4mm] {$Q_{m+1}$}  ;
\draw[dashed]  (7.5,-3)  -- (8,-3) node[right=0mm] {${\nu'}^{\rm{T}}_{m+1}$} ;
\draw[dashed]  (7.5,-3)  -- (7.5,-3.3) ;
\draw[dotted] (7.5,-3.3)  -- (7.5,-3.7) ; 
\draw[dashed]  (7.5,-3.7)  -- (7.5,-4) ;
\draw[dashed]  (7.5,-4)  -- (7,-4)  node[left=-1mm] {$\nu_{m+n}$};
\draw[dashed]  (7.5,-4)  -- (8,-4.5) ; \coordinate (O) at (7.0,-4.3) node at (O) [right=-2mm] {$Q_{m+n}$}  ;
\draw[dashed]  (8,-4.5)  -- (8.5,-4.5) node[right=0mm] {${\nu'}^{\rm{T}}_{m+n}$} ;
\draw[dashed]  (8,-4.5)  -- (8,-5) ;
 \end{tikzpicture}
\end{align}
This  gives the superconformal theories by gluing $N$ chain geometries along the horizontal lines.
The corresponding amplitude is computed by combining the (anti-)vertices,
\be
\ba
\cZ_{\{ \nu_i \} , \{ \nu'_i \} }^{\text{build}}
= 
\sum_{\{ \mu_i \}, \{ \mu'_i \} }
\prod_{i=1}^{m+n} (-Q_i )^{|\mu_i|} (-Q'_i )^{|\mu'_i|} 
& \times
\prod_{i=1}^m C_{{\mu'}^{\rm{T}}_{i-1} \mu_{i} \nu_i}(t,q) C_{{\mu'}_{i} \mu_{i}^{\rm{T}} {\nu'}_i^{\rm{T}}}(q,t) 
\\ &\times
\prod_{i=m+1}^{m+n} \bar{C}_{{\mu'}^{\rm{T}}_{i-1} \mu_{i} \nu_i}(t,q) \bar{C}_{{\mu'}_{i} \mu_{i}^{\rm{T}} {\nu'}_i^{\rm{T}}}(t,q),
\ea
\ee
with the condition $\mu_0 = \mu_{m+n} = \o$ to the top and bottom external legs.%
\footnote{%
Similarly we can consider the partition function of 6d $\mathcal{N} = (1,0)$ gauge theory on $\mathbb{R}^4 \times T^2$ by gluing the top and bottom legs~\cite{HIV}. In that case, we impose the periodic boundary condition, $\mu_0 = \mu_{m+n}.$
}
After some computations,%
\footnote{It might be easier to utilize the operator formalism discussed in \cite{KS} than using the formulae about Schur function.}
we have
\be
\ba
\cZ_{\{ \nu_i \} , \{ \nu'_i \} }^{\text{build}} 
=
&\prod_{i=1}^{m}q^{\f{1}{2}||\nu_i ||^2} t^{\f{1}{2}||{\nu'}^{\rm{T}}_i||^2}\times \prod_{i=m+1}^{m+n}q^{-\f{1}{2}||\nu_i ||^2}  t^{-\f{1}{2}||{\nu'}^{\rm{T}}_i||^2}
\\
&\times
\prod_{i=1}^{m}  \tilde{Z}_{\nu_i}(t,q) \tilde{Z}_{{\nu'}^{{\rm T}}_i}(q,t) \times  \prod_{i=m+1}^{m+n} \tilde{Z}_{\nu_i}(t^{-1},q^{-1}) \tilde{Z}_{{\nu'}^{{\rm T}}_i}(q^{-1},t^{-1})
\\
&\times
\prod_{i,j=1}^\infty
\Biggl[
\prod_{a=1}^m \left(1-Q_a t^{i-\nu_{a,j}^{{\rm T}}-\f{1}{2}} q^{j-{\nu'}_{a,i}-\f{1}{2}} \right)
\prod_{a=m+1}^{m+n} \left(1-Q_a t^{-i+\nu_{a,j}^{{\rm T}}+\f{1}{2}} q^{-j+{\nu'}_{a,i}+\f{1}{2}} \right)
\\
&\times
\prod_{1\leq a < b}^m 
\f{\left(1-Q_{\tau_{a+1,b}} Q'_a t^{i-{\nu'}_{a,j}^{{\rm T}}-\f{1}{2}} q^{j-{\nu}_{b,i}-\f{1}{2}} \right) \left(1-Q_{\tau_{a,b+1}} {Q'}^{-1}_b t^{i-{\nu}_{a,j}^{{\rm T}}-\f{1}{2}} q^{j-{\nu'}_{b,i}-\f{1}{2}} \right)}
{\left(1-Q_{\tau_{a,b}} t^{i-{\nu}_{a,j}^{{\rm T}}} q^{j-{\nu}_{b,i}-1} \right) \left(1-Q_{\tau_{a+1,b+1}} Q'_a {Q'}_b^{-1} t^{i-{\nu'}_{a,j}^{{\rm T}}-1} q^{j-{\nu'}_{b,i}} \right) }
\\
&\times
\prod_{m+1\leq a < b}^{m+n} 
\f{\left(1-Q_{\tau_{a+1,b}} Q'_a t^{-i+{\nu'}_{a,j}^{{\rm T}}+\f{1}{2}} q^{-j+{\nu}_{b,i}+\f{1}{2}} \right) \left(1-Q_{\tau_{a,b+1}} {Q'}^{-1}_b t^{-i+{\nu}_{a,j}^{{\rm T}}+\f{1}{2}} q^{-j+{\nu'}_{b,i}+\f{1}{2}} \right)}
{\left(1-Q_{\tau_{a,b}} t^{-i+{\nu}_{a,j}^{{\rm T}}+1} q^{-j+{\nu}_{b,i}} \right) \left(1-Q_{\tau_{a+1,b+1}} Q'_a {Q'}_b^{-1} t^{-i+{\nu'}_{a,j}^{{\rm T}}} q^{-j+{\nu'}_{b,i}+1} \right) }
 \\
&\times
\prod_{a=1}^m \prod_{b=m+1}^{m+n} 
\f{\left(1-Q_{\tau_{a+1,b}} Q'_a t^{-i-{\nu'}_{a,j}^{{\rm T}}+\f{1}{2}} q^{j+{\nu}_{b,i}-\f{1}{2}} \right)  \left(1-Q_{\tau_{a,b+1}} {Q'}^{-1}_b t^{-i-{\nu}_{a,j}^{{\rm T}}+\f{1}{2}} q^{j+{\nu'}_{b,i}-\f{1}{2}} \right)}
{\left(1-Q_{\tau_{a,b}} t^{-i-{\nu}_{a,j}^{{\rm T}}+1} q^{j+{\nu}_{b,i}-1} \right) \left(1-Q_{\tau_{a+1,b+1}} Q'_a {Q'}_b^{-1} t^{-i-{\nu'}_{a,j}^{{\rm T}}} q^{j+{\nu'}_{b,i}} \right) }
 \Biggr],
\ea
\ee
where we parametrize the K\"ahler parameters as follows:
\begin{subequations}
\begin{align}
 & Q_{\tau_{a,b}} = Q_{\tau_a} Q_{\tau_b}^{-1} \quad (a\leq b), \\
 & Q_{\tau_a} = \prod_{j=a}^{m+n} Q_j Q'_j.
\end{align}
\end{subequations}
Normalizing with the empty configuration $\cZ_{\{ \nu_i = \o \} , \{ \nu'_i = \o \} }^{\text{build}}$, this amplitude gives rise to (a half of) the vector and bifundamental hypermultiplet contributions to the suprgroup gauge theory partition function shown in Sec.~\ref{sec:loc}. The explicit form of the normalized amplitude is
\be\ba
\f{\cZ_{\{ \nu_i \} , \{ \nu'_i \} }^{\text{build}} }{\cZ_{\{ \nu_i = \o \} , \{ \nu'_i = \o \} }^{\text{build}}}
=
&\prod_{i=1}^{m}q^{\f{1}{2}||\nu_i ||^2} t^{\f{1}{2}||{\nu'}^{\rm{T}}_i||^2}\times \prod_{i=m+1}^{m+n}q^{-\f{1}{2}||\nu_i ||^2}  t^{-\f{1}{2}||{\nu'}^{\rm{T}}_i||^2}
\\ &\times
\prod_{1\leq a \leq b}^m 
Z^{{\rm bifund}}_{++}(\{Q \},\{ Q' \};\{\nu\},\{\nu'\}; a,b; t,q)
Z^{{\rm vect}}_{++}(\{Q \},\{ Q' \};\{\nu\},\{\nu'\}; a,b; t,q)
\\ &\times
\prod_{m+1\leq a \leq b}^{m+n} 
Z^{{\rm bifund}}_{--}(\{Q \},\{ Q' \};\{\nu\},\{\nu'\}; a,b; t,q)
Z^{{\rm vect}}_{--}(\{Q \},\{ Q' \};\{\nu\},\{\nu'\}; a,b; t,q)
\\ &\times
\prod_{a=1}^m \prod_{b=m+1}^{m+n} 
\Biggl[
Z^{{\rm bifund}}_{+-}(\{Q \},\{ Q' \};\{\nu\},\{\nu'\}; a,b; t,q)
Z^{{\rm bifund}}_{-+}(\{Q \},\{ Q' \};\{\nu\},\{\nu'\}; a,b; t,q)
\\ &\hspace{20mm} \times
Z^{{\rm vect}}_{+-}(\{Q \},\{ Q' \};\{\nu\}; a,b; t,q)
Z^{{\rm vect}}_{-+}(\{Q \},\{ Q' \};\{\nu'\}; a,b; t,q)
\Biggr],
\ea\ee
where we define the diagonal parts of the building blocks,
\begin{subequations}
\begin{align}
&Z^{{\rm bifund}}_{++}(\{Q \},\{ Q' \};\{\nu\},\{\nu'\}; a,b; t,q)
\nonumber \\
=&
\prod_{(i,j)\in\nu_a}\left(1-Q_{\tau_{a,b+1}} {Q'}^{-1}_b t^{-i+{\nu'}_{b,j}^{{\rm T}}+\f{1}{2}} q^{-j+{\nu}_{a,i}+\f{1}{2}} \right)
\prod_{(i,j)\in \nu'_b}\left(1-Q_{\tau_{a,b+1}} {Q'}^{-1}_b t^{i-{\nu}_{a,j}^{{\rm T}}-\f{1}{2}} q^{j-{\nu'}_{b,i}-\f{1}{2}} \right)
\nonumber \\
&\times 
\prod_{(i,j)\in\nu_b}\left(1-Q_{\tau_{a+1,b}} Q'_a t^{i-{\nu'}_{a,j}^{{\rm T}}-\f{1}{2}} q^{j-{\nu}_{b,i}-\f{1}{2}} \right)^{1-\delta_{a,b}}
\prod_{(i,j)\in \nu'_a}\left(1-Q_{\tau_{a+1,b}} Q'_a t^{-i+{\nu}_{b,j}^{{\rm T}}+\f{1}{2}} q^{-j+{\nu'}_{a,i}+\f{1}{2}} \right)^{1-\delta_{a,b}},
\\
&Z^{{\rm bifund}}_{--}(\{Q \},\{ Q' \};\{\nu\},\{\nu'\}; a,b; t,q)=Z^{{\rm bifund}}_{++}(\{Q \},\{ Q' \};\{\nu\},\{\nu'\}; a,b; t^{-1},q^{-1}),
\\
&Z^{{\rm vect}}_{++}(\{Q \},\{ Q' \};\{\nu\},\{\nu'\};a,b; t,q)
\nonumber  \\
&\qquad
=\prod_{(i,j)\in\nu_a}\f{1}{\left(1-Q_{\tau_{a,b}} t^{-i+{\nu}_{b,j}^{{\rm T}}+1} q^{-j+{\nu}_{a,i}} \right)}
\prod_{(i,j)\in \nu'_a} \f{1}{\left(1-Q_{\tau_{a+1,b+1}} Q'_a {Q'}_b^{-1} t^{-i+{\nu'}_{b,j}^{{\rm T}}} q^{-j+{\nu'}_{a,i}+1} \right) }
\nonumber \\ &\qquad \quad \times
\prod_{(i,j)\in\nu_b}\f{1}{\left(1-Q_{\tau_{a,b}} t^{i-{\nu}_{a,j}^{{\rm T}}} q^{j-{\nu}_{b,i}-1} \right)^{1-\delta_{a,b}}}
\prod_{(i,j)\in \nu'_b} \f{1}{\left(1-Q_{\tau_{a+1,b+1}} Q'_a {Q'}_b^{-1} t^{i-{\nu'}_{a,j}^{{\rm T}}-1} q^{j-{\nu'}_{b,i}} \right)^{1-\delta_{a,b}}},
\\
&Z^{{\rm vect}}_{--}(\{Q \},\{ Q' \};\{\nu\},\{\nu'\} ;a,b;t,q) 
\nonumber  \\
&\qquad
=\prod_{(i,j)\in\nu_a}\f{1}{\left(1-Q_{\tau_{a,b}} t^{i-{\nu}_{b,j}^{{\rm T}}} q^{j-{\nu}_{a,i}-1} \right)}
\prod_{(i,j)\in \nu'_a} \f{1}{\left(1-Q_{\tau_{a+1,b+1}} Q'_a {Q'}_b^{-1} t^{i-{\nu'}_{b,j}^{{\rm T}}-1} q^{j-{\nu'}_{a,i}} \right) }
\nonumber \\ &\qquad \quad \times
\prod_{(i,j)\in\nu_b}\f{1}{\left(1-Q_{\tau_{a,b}} t^{-i+{\nu}_{a,j}^{{\rm T}}+1} q^{-j+{\nu}_{b,i}} \right)^{1-\delta_{a,b}}}
\prod_{(i,j)\in \nu'_b} \f{1}{\left(1-Q_{\tau_{a+1,b+1}} Q'_a {Q'}_b^{-1} t^{-i+{\nu'}_{a,j}^{{\rm T}}} q^{-j+{\nu'}_{b,i}+1} \right)^{1-\delta_{a,b}}},
\end{align}
\end{subequations}
with
\be\ba
\delta_{a,b}=
\begin{cases}
1 \text{ for } a=b,
\\
0 \text{ for } a\neq b,
\end{cases}
\ea\ee
and off-diagonal parts of the building blocks,
\begin{subequations}
\begin{align}
&Z^{{\rm bifund}}_{+-}(\{Q \},\{ Q' \};\{\nu\},\{\nu'\}; a,b; t,q)
\nonumber \\ &\qquad
=\prod_{j=1}^{{\nu}_{a,1}} \prod_{i=1}^{{\nu'}_{b,1}^{\rm{T}}}  \f{\left(1-Q_{\tau_{a,b+1}} {Q'}_b^{-1} t^{-i-{\nu}_{a,j}^{{\rm T}}+\f{1}{2}} q^{j+{\nu'}_{b,i}-\f{1}{2}} \right)}{(1-Q_{\tau_{a,b+1}} {Q'}_b^{-1} t^{-i+\f{1}{2}}q^{j-\f{1}{2}})}
\nonumber \\ &\hspace{10mm}\times
\prod_{(i,j)\in \nu_a}\f{1}{(1-Q_{\tau_{a,b+1}} {Q'}_b^{-1} t^{-{\nu'}^{{\rm T}}_{b,1}-i+\f{1}{2}} q^{j-\f{1}{2}})} \prod_{(i,j)\in \nu'_b} \f{1}{(1-Q_{\tau_{a,b+1}} {Q'}_b^{-1} t^{-i+\f{1}{2}} q^{j+{\nu}_{a,1}-\f{1}{2}})},
\\
&Z^{{\rm bifund}}_{-+}(\{Q \},\{ Q' \};\{\nu\},\{\nu'\} ; a,b; t,q)
\nonumber \\ &\qquad
=\prod_{j=1}^{{\nu'}_{a,1}} \prod_{i=1}^{\nu_{b,1}^{\rm{T}}}  \f{\left(1-Q_{\tau_{a+1,b}} Q'_a t^{-i-{\nu'}_{a,j}^{{\rm T}}+\f{1}{2}} q^{j+{\nu}_{b,i}-\f{1}{2}} \right)}{(1-Q_{\tau_{a+1,b}} Q'_a t^{-i+\f{1}{2}}q^{j-\f{1}{2}})}
\nonumber \\ &\hspace{10mm}\times
\prod_{(i,j)\in \nu'_a}\f{1}{(1-Q_{\tau_{a+1,b}} Q'_a t^{-\nu^{{\rm T}}_{b,1}-i+\f{1}{2}} q^{j-\f{1}{2}})} \prod_{(i,j)\in \nu_b} \f{1}{(1-Q_{\tau_{a+1,b}} Q'_a t^{-i+\f{1}{2}} q^{j+{\nu'}_{a,1}-\f{1}{2}})},
\\
&Z^{{\rm vect}}_{+-}(\{Q \},\{ Q' \};\{\nu'\}; a,b; t,q)
\nonumber \\ &\qquad
=\prod_{j=1}^{{\nu'}_{a,1}} \prod_{i=1}^{{\nu'}_{b,1}^{\rm{T}}}  \f{(1-Q'_a {Q'}^{-1}_b Q_{\tau_{a+1,b+1}} t^{-i}q^{j})}{\left(1-Q'_a {Q'}^{-1}_b Q_{\tau_{a+1,b+1}} t^{-i-{\nu'}_{a,j}^{{\rm T}}} q^{j+{\nu'}_{b,i}} \right)}
\nonumber \\ &\hspace{10mm}\times
\prod_{(i,j)\in \nu'_a} (1-Q'_a {Q'}^{-1}_b Q_{\tau_{a+1,b+1}} t^{-{\nu'}^{{\rm T}}_{b,1}-i} q^{j}) \prod_{(i,j)\in \nu'_b} (1-Q'_a {Q'}^{-1}_b Q_{\tau_{a+1,b+1}} t^{-i} q^{j+{\nu'}_{a,1}}),
\\
&Z^{{\rm vect}}_{-+}(\{Q \},\{ Q' \};\{\nu\}; a,b; t,q)
\nonumber \\ &\qquad
=\prod_{j=1}^{{\nu}_{a,1}} \prod_{i=1}^{{\nu}_{b,1}^{\rm{T}}}  \f{(1-Q_{\tau_{a,b}} t^{-i+1}q^{j-1})}{\left(1-Q_{\tau_{a,b}} t^{-i-{\nu}_{a,j}^{{\rm T}}+1} q^{j+{\nu}_{b,i}-1} \right)}
\nonumber \\ &\hspace{10mm}\times
\prod_{(i,j)\in \nu_a} (1-Q_{\tau_{a,b}} t^{-{\nu}^{{\rm T}}_{b,1}-i+1} q^{j-1}) \prod_{(i,j)\in \nu_b} (1-Q_{\tau_{a,b}} t^{-i+1} q^{j+{\nu}_{a,1}-1}).
\end{align}
\end{subequations}
We remark that the contribution shown here is the instanton part, which is a finite contribution obtained from the full partition function by subtracting the one-loop part.
To show the agreement with the expressions given in Sec.~\ref{sec:loc}, we use following symmetric property under two partitions,
\be\ba
&Z^{{\rm bifund}}_{+-,-+}(\{Q \},\{ Q' \};\{\nu\},\{\nu'\}; a,b; t,q) = Z^{{\rm bifund}}_{+-,-+}(\{Q \},\{ Q' \};\{\nu'\},\{\nu\}, a,b; t,q).
\ea\ee
In this expression, the numerator $\cZ_{\{ \nu_i \} , \{ \nu'_i \} }^{\text{build}}$ is identified with the full partition function, while the denominator $\cZ_{\{ \nu_i = \o \} , \{ \nu'_i = \o \} }^{\text{build}}$ is the one-loop part, so that their ratio gives rise to the instanton partition function by gluing the building blocks.
In addition, the fundamental and antifundamental matter contributions are reproduced by either $\nu_i = \o$ or $\nu'_i = \o$ for $i = 1,\ldots,m+n$.
We can then construct the linear quiver supergroup gauge theory with this building blocks.
For example, $A_k$ quiver gauge theory, $\mathrm{U}(m|n)^{\otimes k}$, is realized with $k+1$ chains (NS5 branes):
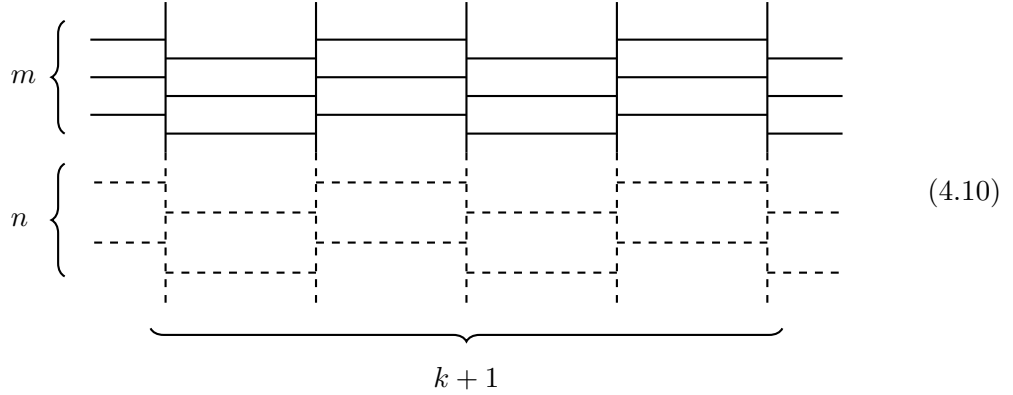
\begin{align}
 \begin{tikzpicture}[thick,baseline=(current bounding box.center)]
  \foreach \x in {0,1,2,3,4}{
  \draw (2*\x,0) -- ++(0,-2);
  \draw[dashed] (2*\x,-2) -- ++(0,-2);
  }
  \draw (0,-.5) -- ++(-1,0);
  \draw (0,-1) -- ++(-1,0);
  \draw (0,-1.5) -- ++(-1,0);
  \draw[dashed] (0,-2.4) -- ++(-1,0);
  \draw[dashed] (0,-3.2) -- ++(-1,0);  
  \foreach \x in {0,4}{
  \draw (\x,-.75) -- ++(2,0);
  \draw (\x,-1.25) -- ++(2,0);
  \draw (\x,-1.75) -- ++(2,0);
  \draw[dashed] (\x,-2.8) -- ++(2,0);
  \draw[dashed] (\x,-3.6) -- ++(2,0);  
  }
  \foreach \x in {2,6}{
  \draw (\x,-.5) -- ++(2,0);
  \draw (\x,-1.) -- ++(2,0);
  \draw (\x,-1.5) -- ++(2,0);
  \draw[dashed] (\x,-2.4) -- ++(2,0);
  \draw[dashed] (\x,-3.2) -- ++(2,0);  
  }
  \draw (8,-.75) -- ++(1,0);
  \draw (8,-1.25) -- ++(1,0);
  \draw (8,-1.75) -- ++(1,0);
  \draw[dashed] (8,-2.8) -- ++(1,0);
  \draw[dashed] (8,-3.6) -- ++(1,0);    
  \draw [decorate,decoration={brace,amplitude=5pt,mirror,raise=4pt},yshift=0pt] (-1.2,-.25) -- ++(0,-1.5) node [right,black,midway,xshift=-1cm,yshift=0cm] {$m$};
 \draw [decorate,decoration={brace,amplitude=5pt,mirror,raise=4pt},yshift=0pt] (-1.2,-2.15) -- ++(0,-1.5) node [right,black,midway,xshift=-1cm,yshift=0cm] {$n$}; 
 \draw [decorate,decoration={brace,amplitude=5pt,mirror,raise=4pt},yshift=0pt] (-.2,-4.2) -- ++(8.4,0) node [black,midway,xshift=0cm,yshift=-.8cm] {$k+1$};  
 \end{tikzpicture}
\end{align}
The corresponding quiver diagram is given by
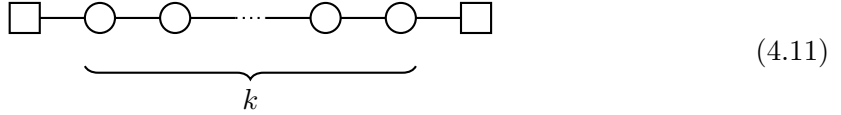
\begin{align}
 \begin{tikzpicture}[thick,baseline=(current bounding box.center)]
  \draw (-1,0) -- (1.8,0);
  \draw [dotted] (1.8,0) -- (2.2,0);
  \draw (2.2,0) -- (5,0);
  \filldraw[fill=white,draw=black] (-1,0)++(-.2,-0.2) rectangle ++(.4,.4);
  \filldraw[fill=white,draw=black] (0,0) circle (.2);
  \filldraw[fill=white,draw=black] (1,0) circle (.2);
  \filldraw[fill=white,draw=black] (3,0) circle (.2);
  \filldraw[fill=white,draw=black] (4,0) circle (.2);
  \filldraw[fill=white,draw=black] (5,0)++(-.2,-0.2) rectangle ++(.4,.4);
  \draw [decorate,decoration={brace,amplitude=5pt,mirror,raise=4pt},yshift=0pt] (-.2,-.5) -- ++(4.4,0) node [below,black,midway,xshift=0cm,yshift=-.3cm] {$k$};
 \end{tikzpicture}
\end{align}
where we denote the gauge node by
\tikz[baseline=-.1cm] \draw[thick] (0,0) circle (.2);
and the flavor node by
\tikz[baseline=.1cm] \draw[thick] (0,0) rectangle ++(.4,.4);, and all the nodes are associated with $\mathrm{U}(m|n)$ group.
In addition to the $A$-type quiver, one can also consider more generic quivers~\cite{Kimura:2019gon}.

\subsection{One-to-Many Correspondence: Gauge Theory and Geometry}

The open string on stack of D-branes gives rise to the non-Abelian gauge field degrees of freedom.
In order to obtain the supergroup gauge field, on the other hand, we need the positive and negative branes~\cite{OT}.
Hence, if considering stack of them, we have to take care of the ordering of the branes.

Firstly let us consider the superconformal theories since we need not take into account the framing factor.
More concretely, here we consider the Hanany--Witten type configuration for $\mathrm{U}(2|1)$ gauge theory with six flavors, which consists of two positive and one negative D4 branes suspended between NS5 branes, and four positive and two negative semi-infinite D4 branes.
In this case, there are three possibilities for changing the position of positive and negative gauge branes as follows: 
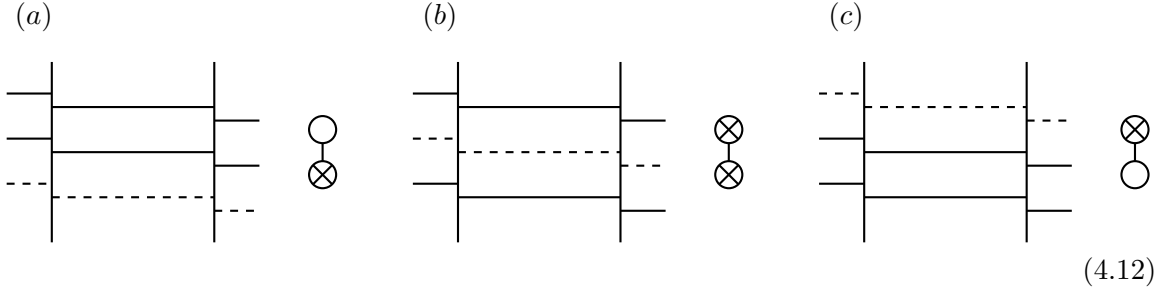
\begin{align}
 \begin{tikzpicture}[thick,baseline=(current bounding box.center),scale=1.2]
  \node at (-.2,2.5) {($a$)};
  \draw (0,0) -- ++(0,2);
  \draw (1.8,0) -- ++(0,2);
  \draw (0,1.5) -- ++(1.8,0);
  \draw (0,1) -- ++(1.8,0);
  \draw[dashed] (0,.5) -- ++(1.8,0);
  \draw (3,1.25) -- ++(0,-.5);
  \draw (-.5,1.65) -- ++(.5,0); 
  \draw (1.8,1.35) -- ++(.5,0); 
  \draw (-.5,1.15) -- ++(.5,0); 
  \draw (1.8,.85) -- ++(.5,0); 
  \draw[dashed] (-.5,.65) -- ++(.5,0);  
  \draw[dashed] (1.8,.35) -- ++(.5,0);  
  \filldraw[fill=white,draw=black] (3,1.25) circle (.15);
  \filldraw[fill=white,draw=black] (3,.75) circle (.15);
   \draw (3,.75)++(135:.15) -- ++(-45:.3);
   \draw (3,.75)++(45:.15) -- ++(-135:.3);     
  \begin{scope}[shift={(4.5,0)}]
   \node at (-.2,2.5) {($b$)};
  \draw (0,0) -- ++(0,2);
  \draw (1.8,0) -- ++(0,2);
  \draw (0,1.5) -- ++(1.8,0);
  \draw[dashed] (0,1) -- ++(1.8,0);
  \draw (0,.5) -- ++(1.8,0);
  \draw (3,1.25) -- ++(0,-.5);
  \draw (-.5,1.65) -- ++(.5,0); 
  \draw (1.8,1.35) -- ++(.5,0); 
  \draw[dashed] (-.5,1.15) -- ++(.5,0); 
  \draw[dashed] (1.8,.85) -- ++(.5,0); 
  \draw (-.5,.65) -- ++(.5,0);  
  \draw (1.8,.35) -- ++(.5,0);  
   \filldraw[fill=white,draw=black] (3,1.25) circle (.15);
   \filldraw[fill=white,draw=black] (3,.75) circle (.15);
   \draw (3,1.25)++(135:.15) -- ++(-45:.3);
   \draw (3,1.25)++(45:.15) -- ++(-135:.3);
   \draw (3,.75)++(135:.15) -- ++(-45:.3);
   \draw (3,.75)++(45:.15) -- ++(-135:.3);     
  \end{scope}
  \begin{scope}[shift={(9,0)}]
   \node at (-.2,2.5) {($c$)};
  \draw (0,0) -- ++(0,2);
  \draw (1.8,0) -- ++(0,2);
  \draw[dashed] (0,1.5) -- ++(1.8,0);
  \draw (0,1) -- ++(1.8,0);
  \draw (0,.5) -- ++(1.8,0);
  \draw (3,1.25) -- ++(0,-.5);
  \draw[dashed] (-.5,1.65) -- ++(.5,0); 
  \draw[dashed] (1.8,1.35) -- ++(.5,0); 
  \draw (-.5,1.15) -- ++(.5,0); 
  \draw (1.8,.85) -- ++(.5,0); 
  \draw (-.5,.65) -- ++(.5,0);  
  \draw (1.8,.35) -- ++(.5,0);  
  \filldraw[fill=white,draw=black] (3,1.25) circle (.15);
  \filldraw[fill=white,draw=black] (3,.75) circle (.15);
   \draw (3,1.25)++(135:.15) -- ++(-45:.3);
   \draw (3,1.25)++(45:.15) -- ++(-135:.3);
  \end{scope}
 \end{tikzpicture}
 \label{eq:HW_U(2|1)}
\end{align}
This one-to-many correspondence between the gauge theory and the brane configurations is a peculiar property to the supergroup theory, which is essentially related to the ambiguity of the simple root decomposition of the supergroup.
Besides the brane configurations, we show the corresponding Dynkin diagrams of $\mathrm{U}(2|1)$:
($a$) \dynkin[root radius=.15cm, edge length=.5cm]{A}{ot}, \
($b$) \dynkin[root radius=.15cm, edge length=.5cm]{A}{tt}, and
($c$) \dynkin[root radius=.15cm, edge length=.5cm]{A}{to},
where the node denoted by \dynkin[root radius=.15cm]{A}{t} is the fermionic node~\cite{Kac:1977em}.
The correspondence is as follows:
We assign the ordinary node to the neighboring pair of D4$^+$-D4$^+$ or D4$^-$-D4$^-$ branes, and the fermionic node is assigned to the neighboring pair of D4$^+$-D4$^-$ branes.
This argument is also applicable to the external flavor branes.

We check the equivalence among these descriptions \eqref{eq:HW_U(2|1)} at the level of the partition function.
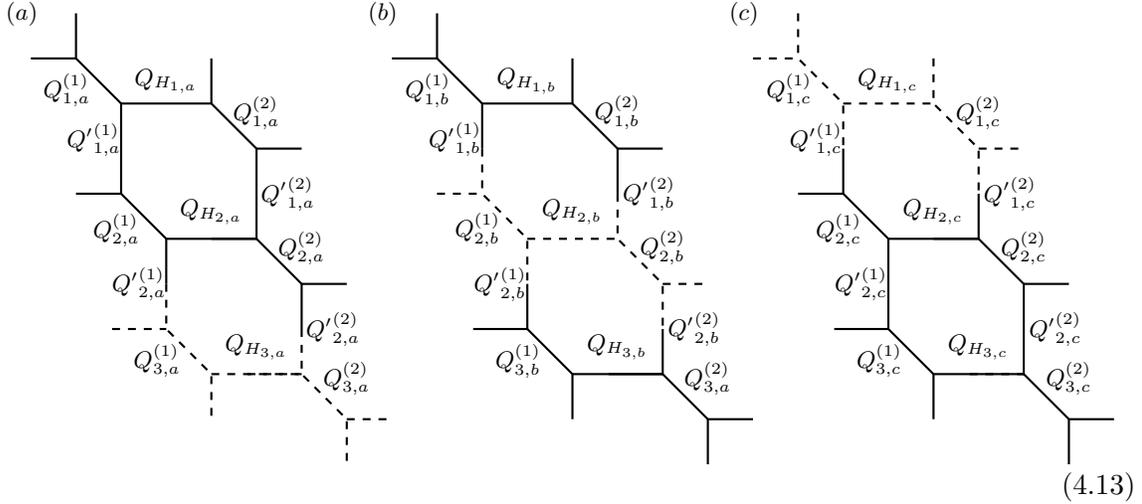
\begin{align}
 \begin{tikzpicture}[thick,baseline=(current bounding box.center),scale=1.2]
 \footnotesize
  \node at (-.6,2) {($a$)};  \node at (1,1.25) {$Q_{H_{1,a}}$} ;  \node at (1.5,-.2) {$Q_{H_{2,a}}$} ;   \node at (2,-1.7) {$Q_{H_{3,a}}$} ;
  \node at (-.1,1.15) {$Q_{1,a}^{(1)}$};  \node at (.2,.6) {${Q'}_{1,a}^{(1)}$};
  \node at (.45,-.4) {$Q_{2,a}^{(1)}$};    \node at (.7,-1) {${Q'}_{2,a}^{(1)}$};
  \node at (.9,-1.85) {$Q_{3,a}^{(1)}$}; 
  \draw (-.5,1.5) -- ++(.5,0);
  \draw (0,2) -- ++(0,-.5) -- ++(.5,-.5) -- ++ (0,-1) -- ++(.5,-.5) -- ++(0,-.5) ;  \draw[dashed] (1,-1.) -- ++(0,-.5) -- ++(.5,-.5) -- ++(0,-.5); 
  \draw (.5,1.0) -- ++(1,0);
  \draw (0,0) -- ++(.5,0);
  \draw (1,-.5) -- ++(1,0);
  \draw[dashed] (.4,-1.5) -- ++(.6,0); 
  \draw[dashed] (1.5,-2) -- ++(1,0);
  \begin{scope}[shift={(1.5,-.5)}]
  \node at (.5,1.4) {$Q_{1,a}^{(2)}$};  \node at (.85,.5) {${Q'}_{1,a}^{(2)}$};
  \node at (1,-.1) {$Q_{2,a}^{(2)}$};    \node at (1.35,-1) {${Q'}_{2,a}^{(2)}$};
  \node at (1.5,-1.55) {$Q_{3,a}^{(2)}$}; 
  \draw (0,2) -- ++(0,-.5) -- ++(.5,-.5) -- ++ (0,-1) -- ++(.5,-.5) -- ++(0,-.5) ;  \draw[dashed] (1,-1.) -- ++(0,-.5) -- ++(.5,-.5) -- ++(0,-.5); 
  \draw (.5,1.0) -- ++(.5,0);
  \draw (0,0) -- ++(.5,0);
  \draw (1,-.5) -- ++(.5,0);
  \draw[dashed] (.4,-1.5) -- ++(.6,0); 
  \draw[dashed] (1.5,-2) -- ++(.5,0);
  \end{scope}
  \begin{scope}[shift={(4,0)}]
  \node at (-.6,2) {($b$)};  \node at (1,1.25) {$Q_{H_{1,b}}$} ;  \node at (1.5,-.2) {$Q_{H_{2,b}}$} ;   \node at (2,-1.7) {$Q_{H_{3,b}}$} ;
  \node at (-.1,1.15) {$Q_{1,b}^{(1)}$};  \node at (.2,.6) {${Q'}_{1,b}^{(1)}$};
  \node at (.45,-.4) {$Q_{2,b}^{(1)}$};    \node at (.7,-1) {${Q'}_{2,b}^{(1)}$};
  \node at (.9,-1.85) {$Q_{3,b}^{(1)}$}; 
  \draw (-.5,1.5) -- ++(.5,0);
  \draw (0,2) -- ++(0,-.5) -- ++(.5,-.5) -- ++ (0,-.5) ; \draw[dashed] (.5,.5) -- ++(0,-.5) -- ++(.5,-.5) -- ++(0,-.5) ;  \draw (1,-1.) -- ++(0,-.5) -- ++(.5,-.5) -- ++(0,-.5); 
  \draw (.5,1.0) -- ++(1,0);
  \draw[dashed] (0,0) -- ++(.5,0);
  \draw[dashed] (1,-.5) -- ++(1,0);
  \draw (.4,-1.5) -- ++(.6,0); 
  \draw (1.5,-2) -- ++(1,0);
  \begin{scope}[shift={(1.5,-.5)}]
  \node at (.5,1.4) {$Q_{1,b}^{(2)}$};  \node at (.85,.5) {${Q'}_{1,b}^{(2)}$};
  \node at (1,-.1) {$Q_{2,b}^{(2)}$};    \node at (1.35,-1) {${Q'}_{2,b}^{(2)}$};
  \node at (1.5,-1.55) {$Q_{3,a}^{(2)}$}; 
  \draw (0,2) -- ++(0,-.5) -- ++(.5,-.5) -- ++ (0,-.5) ; \draw[dashed] (.5,.5) -- ++(0,-.5) -- ++(.5,-.5) -- ++(0,-.5) ;  \draw (1,-1.) -- ++(0,-.5) -- ++(.5,-.5) -- ++(0,-.5); 
  \draw (.5,1.0) -- ++(.5,0);
  \draw[dashed] (1,-.5) -- ++(.5,0);
  \draw (.4,-1.5) -- ++(.6,0); 
  \draw (1.5,-2) -- ++(.5,0);
  \end{scope}
  \end{scope}
  \begin{scope}[shift={(8,0)}]
  \node at (-.6,2) {($c$)};  \node at (1,1.25) {$Q_{H_{1,c}}$} ;  \node at (1.5,-.2) {$Q_{H_{2,c}}$} ;   \node at (2,-1.7) {$Q_{H_{3,c}}$} ;
  \node at (-.1,1.15) {$Q_{1,c}^{(1)}$};  \node at (.2,.6) {${Q'}_{1,c}^{(1)}$};
  \node at (.45,-.4) {$Q_{2,c}^{(1)}$};    \node at (.7,-1) {${Q'}_{2,c}^{(1)}$};
  \node at (.9,-1.85) {$Q_{3,c}^{(1)}$}; 
  \draw[dashed] (-.5,1.5) -- ++(.5,0);
  \draw[dashed] (0,2) -- ++(0,-.5) -- ++(.5,-.5) -- ++ (0,-.5) ;  \draw (.5,.5) -- ++(0,-.5) -- ++ (.5,-.5) -- ++(0,-1) -- ++(.5,-.5) -- ++(0,-.5); 
  \draw[dashed] (.5,1.0) -- ++(1,0);
  \draw (0,0) -- ++(.5,0);
  \draw (1,-.5) -- ++(1,0);
  \draw (.4,-1.5) -- ++(.6,0); 
  \draw (1.5,-2) -- ++(1,0);
  \begin{scope}[shift={(1.5,-.5)}]
  \node at (.5,1.4) {$Q_{1,c}^{(2)}$};  \node at (.85,.5) {${Q'}_{1,c}^{(2)}$};
  \node at (1,-.1) {$Q_{2,c}^{(2)}$};    \node at (1.35,-1) {${Q'}_{2,c}^{(2)}$};
  \node at (1.5,-1.55) {$Q_{3,c}^{(2)}$}; 
  \draw[dashed] (0,2) -- ++(0,-.5) -- ++(.5,-.5) -- ++ (0,-.5) ;  \draw (.5,.5) -- ++(0,-.5) -- ++ (.5,-.5) -- ++(0,-1) -- ++(.5,-.5) -- ++(0,-.5); 
  \draw[dashed] (.5,1.0) -- ++(.5,0);
  \draw (0,0) -- ++(.5,0);
  \draw (1,-.5) -- ++(.5,0);
  \draw[dashed] (.4,-1.5) -- ++(.6,0); 
  \draw (1.5,-2) -- ++(.5,0);
  \end{scope}
  \end{scope}
\normalsize
 \end{tikzpicture}
\end{align}
We take the preferred direction along the horizontal lines, so that the partition functions can be decomposed into left and right building blocks. The equivalence between the gauge theories corresponding to the web diagrams $(a)$ and $(c)$ is trivial because of those shapes, so that here we consider the web diagrams $(a)$ and $(b)$. From the results in Sec.~\ref{BSuCo}, we obtain

\begin{subequations}
\begin{align}
\cZ^{(a)}=
&\sum_{\mu_{1,2,3}}(-Q_{H_{1,a}})^{|\mu_1|} (-Q_{H_{2,a}})^{|\mu_2|} (-Q_{H_{3,a}})^{|\mu_3|} 
 t^{\f{1}{2}(||\mu_1^{\rm T}||^2+||\mu_2^{\rm T}||^2-||\mu_3^{\rm{T}}||^2)}q^{\f{1}{2}(||\mu_1||^2+||\mu_2||^2-||\mu_3||^2)}
 \nonumber \\ &\quad\times
\prod_{1\leq i \leq j}^2 
\Biggl[
Z^{{\rm bifund}}_{++}(\{Q_a^{(1)} \},\{ {Q'}_a^{(1)} \};\{\o\},\{\mu\}; i,j; t,q)
Z^{{\rm vect}}_{++}(\{Q_a^{(1)} \},\{ {Q'}_a^{(1)} \};\{\o\},\{ \mu\}; i,j; t,q)
\nonumber \\ & \qquad\qquad \times
Z^{{\rm bifund}}_{++}(\{Q_a^{(2)} \},\{ {Q'}_a^{(2)} \};\{\mu\},\{\o\}; i,j; t,q)
Z^{{\rm vect}}_{++}(\{Q_a^{(2)} \},\{ {Q'}_a^{(2)} \};\{\mu\},\{ \o\}; i,j; t,q)
\Biggr]
\nonumber \\ & \qquad\qquad \times
Z^{{\rm bifund}}_{--}(\{Q_a^{(1)} \},\{ {Q'}_a^{(1)} \};\{\o\},\{\mu\}; 3,3; t,q)
Z^{{\rm vect}}_{--}(\{Q_a^{(1)} \},\{ {Q'}_a^{(1)} \};\{\o\},\{\mu\}; 3,3; t,q)
\nonumber \\ & \qquad\qquad \times
Z^{{\rm bifund}}_{--}(\{Q_a^{(2)} \},\{ {Q'}_a^{(2)} \};\{\mu\},\{\o\}; 3,3; t,q)
Z^{{\rm vect}}_{--}(\{Q_a^{(2)} \},\{ {Q'}_a^{(2)} \};\{\mu\},\{\o\}; 3,3; t,q)
\nonumber \\ &\times
\prod_{i=1}^2 
\Biggl[
Z^{{\rm bifund}}_{+-}(\{Q_a^{(1)} \},\{ {Q'}_a^{(1)} \};\{\o\},\{ \mu \}; i,3; t,q)
Z^{{\rm bifund}}_{-+}(\{Q_a^{(1)} \},\{ {Q'}_a^{(1)} \};\{\o\},\{\mu\}; i,3; t,q)
\nonumber \\ & \qquad\qquad \times
Z^{{\rm bifund}}_{+-}(\{Q_a^{(2)} \},\{ {Q'}_a^{(2)} \};\{\mu\},\{ \o \}; i,3; t,q)
Z^{{\rm bifund}}_{-+}(\{Q_a^{(2)} \},\{ {Q'}_a^{(2)} \};\{\mu\},\{\o\}; i,3; t,q)
\nonumber \\ & \qquad\qquad \times
Z^{{\rm vect}}_{+-}(\{Q_a^{(1)} \},\{ {Q'}_a^{(1)} \};\{\mu\}; i,3; t,q)
Z^{{\rm vect}}_{-+}(\{Q_a^{(2)} \},\{ {Q'}_a^{(2)} \};\{\mu\},; i,3; t,q)
\Biggr]
\label{PFU(2|1)(a)},
\\
\cZ^{(b)}=
&\sum_{\mu_{1,2,3}}(-Q_{H_{1,b}})^{|\mu_1|} (-Q_{H_{2,b}})^{|\mu_2|} (-Q_{H_{3,b}})^{|\mu_3|} 
 t^{\f{1}{2}(||\mu_1^{\rm T}||^2+||\mu_2^{\rm T}||^2-||\mu_3^{\rm{T}}||^2)}q^{\f{1}{2}(||\mu_1||^2+||\mu_2||^2-||\mu_3||^2)}
 \nonumber \\
&\quad\times
\prod_{1\leq i \leq j,i\neq2, j\neq2}^3 
\Biggl[
Z^{{\rm bifund}}_{++}(\{Q_b^{(1)} \},\{ {Q'}_b^{(1)} \};\{\o\},\{\mu\}; i,j; t,q)
Z^{{\rm vect}}_{++}(\{Q_b^{(1)} \},\{ {Q'}_b^{(1)} \};\{\o\},\{ \mu\}; i,j; t,q)
\nonumber \\ & \qquad\qquad \times
Z^{{\rm bifund}}_{++}(\{Q_b^{(2)} \},\{ {Q'}_b^{(2)} \};\{\mu\},\{\o\}; i,j; t,q)
Z^{{\rm vect}}_{++}(\{Q_b^{(2)} \},\{ {Q'}_b^{(2)} \};\{\mu\},\{ \o\}; i,j; t,q)
\Biggr]
\nonumber \\ & \qquad\qquad \times
Z^{{\rm bifund}}_{--}(\{Q_b^{(1)} \},\{ {Q'}_b^{(1)} \};\{\o\},\{\mu\}; 2,2; t,q)
Z^{{\rm vect}}_{--}(\{Q_b^{(1)} \},\{ {Q'}_b^{(1)} \};\{\o\},\{\mu\}; 2,2; t,q)
\nonumber \\ & \qquad\qquad \times
Z^{{\rm bifund}}_{--}(\{Q_b^{(2)} \},\{ {Q'}_b^{(2)} \};\{\mu\},\{\o\}; 2,2; t,q)
Z^{{\rm vect}}_{--}(\{Q_b^{(2)} \},\{ {Q'}_b^{(2)} \};\{\mu\},\{\o\}; 2,2; t,q)
\nonumber \\ &\times
\Biggl[
Z^{{\rm bifund}}_{+-}(\{Q_b^{(1)} \},\{ {Q'}_b^{(1)} \};\{\o\},\{ \mu \}; 1,2; t,q)
Z^{{\rm bifund}}_{-+}(\{Q_b^{(1)} \},\{ {Q'}_b^{(1)} \};\{\o\},\{\mu\}; 1,2, t,q)
\nonumber \\ & \quad \times
Z^{{\rm bifund}}_{+-}(\{Q_b^{(1)} \},\{ {Q'}_b^{(1)} \};\{\o\},\{ \mu \}; 2,3; t^{-1},q^{-1})
Z^{{\rm bifund}}_{-+}(\{Q_b^{(1)} \},\{ {Q'}_b^{(1)} \};\{\o\},\{\mu\}; 2,3, t^{-1},q^{-1})
\nonumber \\ & \quad \times
Z^{{\rm bifund}}_{+-}(\{Q_b^{(2)} \},\{ {Q'}_b^{(2)} \};\{\mu\},\{ \o \}; 1,2; t,q)
Z^{{\rm bifund}}_{-+}(\{Q_b^{(2)} \},\{ {Q'}_b^{(2)} \};\{\mu\},\{\o\}; 1,2; t,q)
\nonumber \\ & \quad \times
Z^{{\rm bifund}}_{+-}(\{Q_b^{(2)} \},\{ {Q'}_b^{(2)} \};\{\mu\},\{ \o \}; 2,3; t^{-1},q^{-1})
Z^{{\rm bifund}}_{-+}(\{Q_b^{(2)} \},\{ {Q'}_b^{(2)} \};\{\mu\},\{\o\}; 2,3; t^{-1},q^{-1})
\nonumber \\ & \quad \times
Z^{{\rm vect}}_{+-}(\{Q_b^{(1)} \},\{ {Q'}_b^{(1)} \};\{\mu\}; 1,2; t,q)
\check{Z}^{{\rm vect}}_{+-}(\{Q_b^{(1)} \},\{ {Q'}_b^{(1)} \};\{\mu\};  2,3; t^{-1},q^{-1})
\nonumber \\ & \quad \times
Z^{{\rm vect}}_{-+}(\{Q_b^{(2)} \},\{ {Q'}_b^{(2)} \};\{\mu\}; 1,2; t,q)
\check{Z}^{{\rm vect}}_{-+}(\{Q_b^{(2)} \},\{ {Q'}_b^{(2)} \};\{\mu\};  2,3; t^{-1},q^{-1})
\Biggr],
\end{align} 
\end{subequations}
where $\check{Z}^{{\rm vect}}_{+-,-+}$ is the shifted contribution by $(t/q)^{\pm 1}$,
\begin{subequations}
\begin{align}
&\check{Z}^{{\rm vect}}_{+-}(\{Q \},\{ Q' \};\{\nu\}; a,b; t,q)
\nonumber \\ &\qquad
=\prod_{j=1}^{{\nu}_{a,1}} \prod_{i=1}^{{\nu}_{b,1}^{\rm{T}}}  \f{(1- Q_{\tau_{a,b}} t^{-i}q^{j})}{\left(1-Q_{\tau_{a,b}} t^{-i-{\nu}_{a,j}^{{\rm T}}} q^{j+{\nu}_{b,i}} \right)}
\prod_{(i,j)\in \nu_a} (1- Q_{\tau_{a,b}} t^{-{\nu}^{{\rm T}}_{b,1}-i} q^{j}) \prod_{(i,j)\in \nu_b} (1- Q_{\tau_{a,b}} t^{-i} q^{j+{\nu}_{a,1}}),
\\
&\check{Z}^{{\rm vect}}_{-+}(\{Q \},\{ Q' \};\{\nu'\}; a,b; t,q)
\nonumber \\ &\qquad
=\prod_{j=1}^{{\nu'}_{a,1}} \prod_{i=1}^{{\nu'}_{b,1}^{\rm{T}}}  \f{(1-Q'_a {Q'}^{-1}_b Q_{\tau_{a+1,b+1}} t^{-i+1}q^{j-1})}{\left(1- Q'_a {Q'}^{-1}_b Q_{\tau_{a+1,b+1}} t^{-i-{\nu'}_{a,j}^{{\rm T}}+1} q^{j+{\nu'}_{b,i}-1} \right)}
\nonumber \\ &\hspace{10mm}\times
\prod_{(i,j)\in \nu'_a} (1- Q'_a {Q'}^{-1}_b Q_{\tau_{a+1,b+1}} t^{-{\nu'}^{{\rm T}}_{b,1}-i+1} q^{j-1}) \prod_{(i,j)\in \nu'_b} (1-Q'_a {Q'}^{-1}_b Q_{\tau_{a+1,b+1}} t^{-i+1} q^{j+{\nu'}_{a,1}-1}).
\end{align}
\end{subequations}
We find the agreement between these partition functions under the following parameter correspondence,
\begin{subequations}
\begin{align}
&Q_{1,a}^{(I)} = Q_{1,b}^{(I)},~{Q'}_{1,a}^{(I)} =Q^{(I)}_{\tau_{2,a}} {Q'}_{1,b}^{(I)},~
Q_{2,a}^{(I)}=Q_{3,b}^{(I)},~
{Q'}_{2,a}^{(I)} = \left( Q^{(I)}_{2,b} {Q'}^{(I)}_{2,b} Q^{(I)}_{3,b} \right)^{-1},~
Q_{3,a}^{(I)}=Q_{2,b}^{(I)},
\end{align}
\end{subequations}
where $I=1,2$.
To reach the results, we flip the K\"ahler parameters, e.g.,
\be\ba
&\prod_{j=1}^{\mu_{3,1}^{\rm T}} \prod_{i=1}^{\mu_{2,1}} \f{1-Q_2 t^i q^{-j}}{1-Q_2 t^{i+\mu_{3,j}^{\rm T}} q^{-j-\mu_{2,i}}}
 \prod_{(i,j)\in \mu_3} \left(1-Q_2 t^{\mu^{\rm T}_{2,1}+i} q^{-j} \right) \prod_{(i,j) \in \mu_2} \left( 1 - Q_2 t^{i} q^{-j-\mu_{3,1}} \right)
 \\
& =
 Q_2^{|\mu_2|+|\mu_3|} \tilde{f}_{\mu_2}(t,q)  \tilde{f}_{\mu_3}(t,q)
 \\
&\qquad\times
\prod_{j=1}^{\mu_{3,1}^{\rm T}} \prod_{i=1}^{\mu_{2,1}} \f{1-Q^{-1}_2 t^{-i} q^{j}}{1-Q_2 t^{-i-\mu_{3,j}^{\rm T}} q^{j+\mu_{2,i}}}
 \prod_{(i,j)\in \mu_3} \left(1-Q_2^{-1} t^{-\mu^{\rm T}_{2,1}-i} q^{j} \right) \prod_{(i,j) \in \mu_2} \left( 1 - Q_2 t^{-i} q^{j+\mu_{3,1}} \right).
\ea\ee

In addition to above ambiguities, we also can change the position of flavor brane, which leads to additional ambiguity. For example, we consider following brane set up and corresponding $(p,q)$ 5-brane web diagram as following:
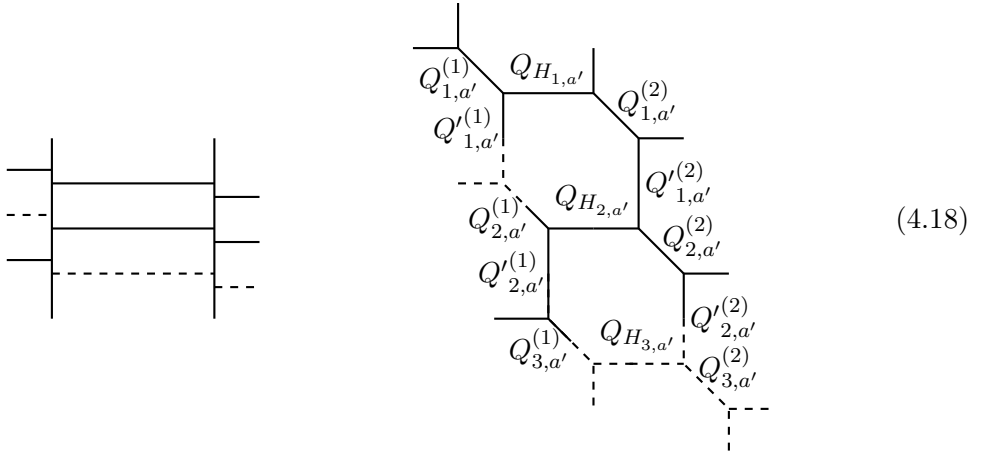
\begin{align}
 \begin{tikzpicture}[thick,baseline=(current bounding box.center),scale=1.2]
  \draw (0,0) -- ++(0,2);
  \draw (1.8,0) -- ++(0,2);
  \draw (0,1.5) -- ++(1.8,0);
  \draw (0,1) -- ++(1.8,0);
  \draw[dashed] (0,.5) -- ++(1.8,0);
  \draw (-.5,1.65) -- ++(.5,0); 
  \draw (1.8,1.35) -- ++(.5,0); 
  \draw[dashed] (-.5,1.15) -- ++(.5,0); 
  \draw (1.8,.85) -- ++(.5,0); 
  \draw (-.5,.65) -- ++(.5,0);  
  \draw[dashed] (1.8,.35) -- ++(.5,0);  
  \begin{scope}[shift={(4.5,1.5)}]
  \node at (1,1.25) {$Q_{H_{1,a'}}$} ;  \node at (1.5,-.2) {$Q_{H_{2,a'}}$} ;   \node at (2,-1.7) {$Q_{H_{3,a'}}$} ;
  \node at (-.1,1.15) {$Q_{1,a'}^{(1)}$};  \node at (.1,.6) {${Q'}_{1,a'}^{(1)}$};   
  \node at (.45,-.4) {$Q_{2,a'}^{(1)}$};    \node at (.6,-1) {${Q'}_{2,a'}^{(1)}$}; 
  \node at (.9,-1.85) {$Q_{3,a'}^{(1)}$}; 
  \draw (-.5,1.5) -- ++(.5,0);
  \draw (0,2) -- ++(0,-.5) -- ++(.5,-.5) -- ++ (0,-.5); \draw[dashed] (.5,.5) --++(0,-.5)-- ++(.25,-.25) ; 
  \draw (.75,.-.25) --++ (.25,-.25) --++ (0,-1)--++(.25,-.25);  \draw[dashed] (1,-1.) -- ++(0,-.5) -- ++(.5,-.5) -- ++(0,-.5); 
  \draw (.5,1.0) -- ++(1,0);
  \draw[dashed] (0,0) -- ++(.5,0);
  \draw (1,-.5) -- ++(.5,0);
  \draw (.4,-1.5) -- ++(.6,0); 
  \draw[dashed] (1.5,-2) -- ++(.5,0);
  \begin{scope}[shift={(1.5,-.5)}]
  \node at (.6,1.4) {$Q_{1,a'}^{(2)}$};  \node at (.95,.5) {${Q'}_{1,a'}^{(2)}$};
  \node at (1.1,-.1) {$Q_{2,a'}^{(2)}$};    \node at (1.45,-1) {${Q'}_{2,a'}^{(2)}$};
  \node at (1.5,-1.55) {$Q_{3,a'}^{(2)}$}; 
  \draw (0,2) -- ++(0,-.5) -- ++(.5,-.5) -- ++ (0,-1) -- ++(.5,-.5) -- ++(0,-.5) ;  \draw[dashed] (1,-1.) -- ++(0,-.5) -- ++(.5,-.5) -- ++(0,-.5); 
  \draw (.5,1.0) -- ++(.5,0);
  \draw (0,0) -- ++(.5,0);
  \draw (1,-.5) -- ++(.5,0);
  \draw[dashed] (.4,-1.5) -- ++(.6,0); 
  \draw[dashed] (1.5,-2) -- ++(.5,0);
\end{scope}
\end{scope}
   \end{tikzpicture}
    \label{WebAmbFlav}
\end{align}
We do not repeat the same computation here, but one can show that by setting the K\"ahler parameters to
\be\ba
&Q^{(1)}_{1,a'}=Q^{(1)}_{1,a},~
\\
&{Q'}^{(1)}_{1,a'}={Q'}^{(1)}_{1,a} {Q}^{(1)}_{2,a} {Q'}^{(1)}_{2,a} ,~
{Q}^{(1)}_{2,a'}=\left( {Q'}^{(1)}_{2,a} \right)^{-1},~
{Q'}^{(1)}_{2,a'}=\left( {Q}^{(1)}_{2,a} \right)^{-1},~
{Q}^{(1)}_{3,a'}={Q}^{(1)}_{2,a} {Q'}^{(1)}_{2,a} {Q}^{(1)}_{3,a} ,~
\\
&Q_{H_{2,a'}} ={Q}^{(1)}_{2,a} {Q'}^{(1)}_{2,a} Q_{H_{2,a}},
\ea\ee
the partition function of \eqref{WebAmbFlav}  agrees with \eqref{PFU(2|1)(a)}.

Therefore, we conclude that there are non-unique web diagram descriptions for a supergroup gauge theory, however, all of the descriptions are equivalent in the sense that they provide the the partition functions under the suitable parameter correspondence.


\section{Towards Geometric Engineering for Pure Gauge Theories}\label{sec:chain_gmtry}
\subsection{Building block}
To consider more general supergroup gauge theories, especially the pure supergroup gauge theories, we calculate another kinds of chain geometry, constructed by both the ordinary vertex and the anti-vertex given by the following web diagrams with non-trivial boundary conditions along the external lines:
\begin{align}
\begin{tikzpicture}[thick,baseline=(current bounding box.center)]
\footnotesize
\coordinate (O) at (-2.5,0.4) node at (O) [left=8mm] {$\cZ_{\{ \nu_i \} }^L ~=~$}  ;
\draw[] (1,1) -- (1,1.5) ;
\draw[] (1,1) -- (1.5,1) node[right] {$\nu_1$};
\draw[] (0.5,0.5) -- (1,1) ;  \coordinate (O) at (0.8,0.8) node at (O) [left=-0mm] {$Q_1$}  ;
\draw[] (0.5,0.5) -- (1,0.5) node[right=0mm] {$\nu_2$} ;
\draw[] (0.2,0.35) -- (0.5,0.5) ; 
\draw[dotted] (0.2,0.35) -- (-0.1,0.15) ; 
\draw[] (-0.1,0.15) -- (-0.3,0.05) ; 
\draw[]  (-0.3,0.05) -- (0.3,0.05) node[right=0mm] {$\nu_m$} ; 
\draw[]  (-0.3,0.05) -- (-0.5,-0.01) ;
\draw[dashed]  (-0.5,-0.01) -- (-1.4,-0.31) ; \coordinate (O) at (-0.5,0.1) node at (O) [left=--1mm] {$Q_m$}  ;
\draw[dashed]   (-1.4,-0.31)  -- (-0.6,-0.31)  node[right=0mm] {$\nu_{m+1}$}  ; 
\draw[dashed]  (-1.4,-0.31) -- (-2.0,-0.46) ;
\draw[dotted]  (-2.0,-0.46) -- (-2.2,-0.51) ;
\draw[dashed] (-2.2,-0.51) -- (-2.8,-0.66) ;
\draw[dashed]  (-2.8,-0.66)  -- (-2.2,-0.66) ; \coordinate (O) at (-2.2,-0.7) node at (O) [right=-1mm] {$\nu_{m+n}$}  ;
\draw[dashed]  (-2.8,-0.66)  -- (-3.3,-0.76) ;
\coordinate (O) at (5.5,0.4) node at (O) [left=10mm] {$\cZ_{\{ \nu_i \} }^R ~=~$}  ;
\draw[] (5.5,1) -- (5.5,1.5) ;
\draw[] (5,1) -- (5.5,1) node[left=4mm] {$\nu^{\rm{T}}_1$};
\draw[] (6,0.5) -- (5.5,1) ;  \coordinate (O) at (5.8,0.9) node at (O) [right=-1mm] {$Q_1$}  ;
\draw[] (6,0.5) -- (5.5,0.5) node[left =0mm] {$\nu^{\rm{T}}_2$} ;
\draw[] (6.2,0.4) -- (6,0.5) ; 
\draw[dotted] (6.2,0.4) -- (6.6,0.2) ; 
\draw[] (6.6,0.2) -- (6.8,0.1) ; 
\draw[]  (6.3,0.1) -- (6.8,0.1) node[left=4mm] {$\nu^{\rm{T}}_m$} ; 
\draw[]  (6.8,0.1) -- (7.4,-0.1) ;
\draw[dashed]  (7.4,-0.1) -- (8,-0.3) ; \coordinate (O) at (8,0.1) node at (O) [left=--1mm] {$Q_m$}  ;
\draw[dashed]   (8,-0.3)  -- (7.5,-0.3)  node[left=0mm] {$\nu^{\rm{T}}_{m+1}$}  ; 
\draw[dashed]  (8,-0.3) -- (8.8,-0.5) ;
\draw[dotted]  (8.8,-0.5) -- (9.2,-0.6) ;
\draw[dashed] (9.2,-0.6) -- (9.6,-0.7) ;
\draw[dashed]  (9.6,-0.7)  -- (9.1,-0.7) ; \coordinate (O) at (8,-0.7) node at (O) [right=2mm] {$\nu^{\rm{T}}_{m+n}$}  ;
\draw[dashed]  (9.6,-0.7)  -- (10.1,-0.8) ;
\end{tikzpicture}
 \label{eq:chain_geom}
\end{align}
By gluing them along the horizontal lines, we can construct the pure $\mathrm{U}(m|n)$ supergroup gauge theories, which is a natural generalization of the $\mathrm{U}(n)$ gauge theory~\cite{Iqbal:2003zz,Eguchi:2003sj}.
The chain geometry amplitude is written using the vertices,
\be
\ba
\cZ_{\{ \nu_i \} }^L
= 
\sum_{\{ \mu_i \}, \{\eta_j \}} 
&\prod_{i=1}^{m+n} (-Q_i )^{|\mu_i|} 
\times \prod_{i=1}^{m+m} f^{-1}_{\mu_i}(q,t)
\\ 
&\times
\prod_{i=1}^m C_{\mu_i \lambda_{i-1}^{\rm{T}} \nu_i}(t,q) \times \prod_{i=m+1}^{m+n} \bar{C}_{\mu_i \lambda_{i-1}^{\rm{T}} \nu_i}(t,q),
\ea
\ee
where we impose $\lambda_0=\mu_{m+n}=\o$ corresponding to the most top and bottom external lines. Applying the formulae for the skew-Schur function, we find
\be
\ba
\cZ_{\{ \nu_i \} }^L
=
& \prod_{i=1}^{m} q^{\f{1}{2}||\nu_i||^2}  \times \prod_{i=m+1}^{m+n} q^{-\f{1}{2}||\nu_i||^2}  \times  \prod_{i=1}^{m}  \tilde{Z}_{\nu_i}(t,q) \times  \prod_{i=m+1}^{m+n} \tilde{Z}_{\nu_i}(t^{-1},q^{-1})
\\ &\times
\prod_{i,j=1}^\infty \Biggl[
\prod_{1\leq a< b}^m \f{1}{(1-Q_{ab} t^{i-\nu_{b,j}^{\rm{T}}}q^{j-\nu_{a,i}-1})}
\times
\prod_{m+1\leq a< b}^{m+n} \f{1}{(1-Q_{ab} t^{-i+\nu_{b,j}^{\rm{T}}+1}q^{-j+\nu_{a,i}})}
\\ &\times
\prod_{a=1}^{m} \prod_{b=m+1}^{m+n} \f{1}{(1-Q_{ab} t^{i+\nu_{b,j}^{\rm{T}}}q^{-j-\nu_{a,i}})}
\Biggr],
\ea
\ee
where
\be
\ba
Q_{ab}= \prod_{j=a}^{b-1} Q_j . 
\ea
\ee
Similarly, we find the expression for the other case, $\cZ_{\{ \nu_i \} }^R$,
\be
\ba
\cZ_{\{ \nu_i \} }^R
=
& \prod_{i=1}^{m} t^{\f{1}{2}||\nu^{\rm{T}}_i||^2}\times  \prod_{i=m+1}^{m+n} t^{-\f{1}{2}||\nu_i^{\rm T}||^2} \times  \prod_{i=1}^{m} \tilde{Z}_{\nu^{\rm{T}}_i}(q,t)  \times \prod_{i=m+1}^{m+n} \tilde{Z}_{\nu^{\rm{T}}_i}(q^{-1},t^{-1})
\\ &\times
\prod_{i,j=1}^\infty \Biggl[
\prod_{1\leq a< b}^m \f{1}{(1-Q_{ab} t^{i-\nu_{b,j}^{\rm{T}}-1}q^{j-\nu_{a,i}})}
\times
\prod_{m+1\leq a< b}^{m+n} \f{1}{(1-Q_{ab} t^{-i+\nu_{b,j}^{\rm{T}}}q^{-j+\nu_{a,i}+1})}
\\ &\times
\prod_{a=1}^{m} \prod_{b=m+1}^{m+n} \f{1}{(1- Q_{ab} t^{i+\nu_{b,j}^{\rm{T}}+1}q^{-j-\nu_{a,i}-1})}
\Biggr].
\ea
\ee
The product of these building block with a normalization factor,
\be
\ba
\cZ^{\text{vect}}_{\{\nu_i\}} = \frac{\cZ_{\{ \nu_i \} }^L \cZ_{\{ \nu_i \} }^R}{\cZ_{\{ \nu_i =\o\} }^L \cZ_{\{ \nu_i = \o \} }^R},
\label{NV}
\ea
\ee
agrees with the instanton partition function of 5d $\mathcal{N}=1$ $\mathrm{U}(m|n)$ pure gauge theory obtained in Sec.~\ref{sec:loc}.
The extra factor coming from the framing is identified with the Chern--Simons term in 5d gauge theory.

Here we provide each normalized building blocks, which can be given by $Z^{\text{vect}}_{++,+-,-+,--}$,
\begin{subequations}
\begin{align}
\frac{\cZ_{\{ \nu_i \} }^L}{\cZ_{\{ \nu_i =\o\} }^L}
&= \prod_{i=1}^{m} q^{\f{1}{2}||\nu_i||^2}  \times \prod_{i=m+1}^{m+n} q^{-\f{1}{2}||\nu_i||^2}
\nonumber \\
&\qquad \times
\Biggl[
\prod_{1\leq a\leq b}^m Z^{\text{vect}}_{++}(\{ Q \} ;\{ \o \},\{ \nu^{\rm T} \};a,b,q,t)\Big|_{\text{Rep}}
\times 
\prod_{m+1\leq a\leq b}^{m+n} Z^{\text{vect}}_{--}(\{ Q \};\{ \o \},\{ \nu^{\rm T} \};a,b,q,t)\Big|_{\text{Rep}}
\nonumber \\
&\qquad \times
\prod_{a=1}^m \prod_{b=m+1}^{m+n}Z^{\text{vect}}_{+-}(\{ Q \};\{ \o \},\{ \nu^{\rm T} \};a,b,q,t)\Big|_{\text{Rep}}
\Biggr] ,
\label{LeftPure}
\\
 \frac{\cZ_{\{ \nu_i \} }^R}{\cZ_{\{ \nu_i = \o \} }^R}
&= \prod_{i=1}^{m} t^{\f{1}{2}||\nu^{\rm{T}}_i||^2}\times  \prod_{i=m+1}^{m+n} t^{-\f{1}{2}||\nu_i^{\rm T}||^2}
\nonumber \\
&\qquad \times
\Biggl[
\prod_{1\leq a\leq b}^m  Z^{\text{vect}}_{++}(\{ Q \};\{ \nu^{\rm T} \},\{ \o \};a,b,q,t)\Big|_{\text{Rep}}
\times 
\prod_{m+1\leq a\leq b}^{m+n} Z^{\text{vect}}_{--}(\{ Q \};\{ \nu^{\rm T} \},\{ \o \};a,b,q,t)\Big|_{\text{Rep}}
\nonumber \\
&\qquad \times
\prod_{a=1}^m \prod_{b=m+1}^{m+n} Z^{\text{vect}}_{-+}(\{ Q \};\{ \nu^{\rm T} \},\{ \o \};a,b,q,t)\Big|_{\text{Rep}}
\Biggr],
\end{align}
\end{subequations}
where ``Rep'' denotes the following replacement,
\be\ba
Q_{\tau_{a,b}}\to Q_{ab},~Q'_a {Q'}^{-1}_b Q_{\tau_{a+1,b+1}} \to Q_{ab}.
\ea\ee

We remark the relation between the web diagram discussed above and the brane setup~\cite{DHJV,KP1}.
It is known that the web diagram is interpreted as the web of $(p,q)$ five branes, and the current setup, associated with 5d $\mathcal{N} = 1$ $\mathrm{U}({m|n})$ gauge theory, is T-dual to the Hanany--Witten type configuration for 4d $\mathcal{N} = 2$ theory:
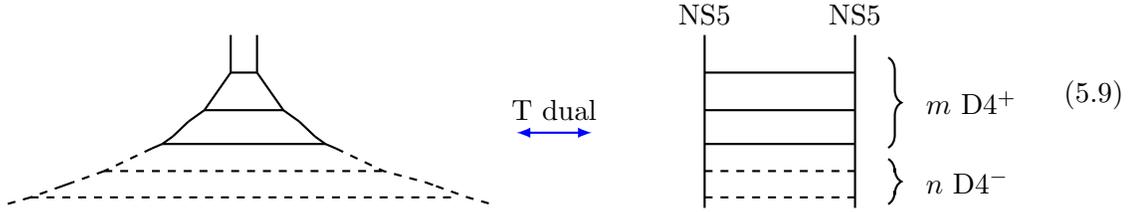
\begin{align}
\begin{tikzpicture}[thick,baseline=(current bounding box.center)]
\begin{scope}[xscale=.7]
 \draw[] (1,1) -- (1,1.5) ;
\draw[] (0.5,0.5) -- (1,1) ; 
\draw[] (0.2,0.35) -- (0.5,0.5) ; 
\draw[] (0.2,0.35) -- (-0.1,0.15) ; 
\draw[] (-0.1,0.15) -- (-0.3,0.05) ; 
\draw[]  (-0.3,0.05) -- (-0.5,-0.01) ;
\draw[dashed]  (-0.5,-0.01) -- (-1.4,-0.31) ; 
\draw[dashed]  (-1.4,-0.31) -- (-2.0,-0.46) ;
\draw[]  (-2.0,-0.46) -- (-2.2,-0.51) ;
\draw[dashed] (-2.2,-0.51) -- (-2.8,-0.66) ;
\draw[dashed]  (-2.8,-0.66)  -- (-3.3,-0.76) ;
\draw[] (1.5,1.5) -- (1.5,1) -- (2,.5) -- ++(.3,-0.15);
\draw[] (2.3,.35) -- ++(.3,-.2);
\draw[] (2.6,.15) -- (2.8,.05) -- ++(.2,-.06);
\draw[dashed] (3,-.01) -- ++(.9,-.3) -- ++(.8,-.15);
\draw[dashed] (4.7,-.46) -- ++(.2,-.05);
\draw[dashed] (4.9,-.51) -- ++(.6,-.15) -- ++(.5,-.1);
\draw[] (1,1) -- (1.5,1);    
\draw[] (0.5,0.5) -- (2,0.5) ; 
\draw[]  (-0.3,0.05) -- (2.8,0.05) ; 
\draw[dashed]   (-1.4,-0.31)  -- ++(5.3,0) ;
\draw[dashed]  (-2.8,-0.66)  -- ++(8.1,0) ;
\end{scope}
   \draw[latex-latex,blue] (4.5,.2) -- ++(1,0);
   \node at (5,.5) {T dual};
   \begin{scope}[shift={(7,0)}]
    \draw (0,1.5) node [above] {NS5} -- ++(0,-2.3);
    \draw (2,1.5) node [above] {NS5} -- ++(0,-2.3);
    \draw (0,1) -- ++(2,0);
    \draw (0,.5) -- ++(2,0);    
    \draw (0,.05) -- ++(2,0);
    \draw[dashed] (0,-.31) -- ++(2,0);
    \draw[dashed] (0,-.66) -- ++(2,0);
    \draw [decorate,decoration={brace,amplitude=5pt,raise=4pt},yshift=0pt] (2.3,1.2) -- (2.3,0) node [right,black,midway,xshift=.5cm] {$m$ D4$^+$};  
     \draw [decorate,decoration={brace,amplitude=5pt,raise=4pt},yshift=0pt] (2.3,-.15) -- (2.3,-.75) node [right,black,midway,xshift=.5cm] {$n$ D4$^-$};    \end{scope}
  \end{tikzpicture}
  \label{U(m|n)}
\end{align}
Under this duality, D5 branes are converted to D4 branes in Type IIA theory.
The horizontal solid line stands for the positive D4 branes and the dashed line is the negative D4 branes, denoted by D4$^+$ and D4$^-$, respectively.

\subsection{Vertex vs. Anti-Vertex}
\label{equiv}
We discuss a possible relation between the ordinary vertex and the anti-vertex.
Here we consider the unrefined case $t = q$ for simplicity,
\be\ba
\bar{C}_{\lambda \mu \nu}(q) =\bar{C}_{\lambda \mu \nu}(q,q)
=
q^{\f{1}{2}(\kappa_\mu+\kappa_\nu)} s_{\nu^{\rm T}}(q^\rho) \sum_{\eta} s_{\lambda^{\rm T}/\eta}(q^{\rho+\nu}) s_{\mu/\eta}(q^{\rho+\nu^{\rm T}}) .
\ea\ee
One can show that the anti-vertex is equivalent to the ordinary vertex up to the framing factor,
\be
\ba
C_{\lambda\mu\nu}(q) =f_{\lambda} f_{\mu}^{-1} f_{\nu}  \bar{C}_{\mu\lambda\nu}(q),
\label{eq:vertex_anti-vertex}
\ea
\ee
where $f_\mu  = f_\mu (q,q)$. 
Graphically this relation can be expressed as follows:
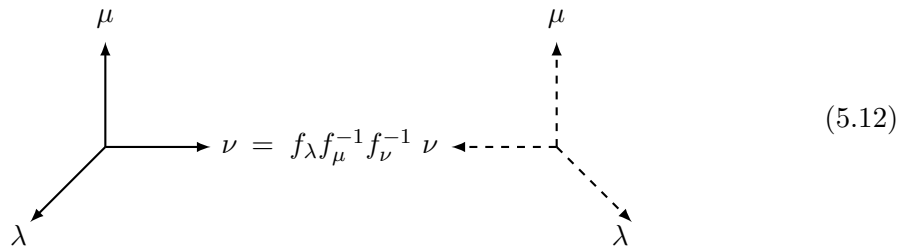
\begin{align}
 \begin{tikzpicture}[thick,baseline=(current bounding box.center)]
 \draw[-latex] (1-6,1) -- (0-6,0) node[below left = -1mm] {$\lambda$}  ;
\draw[-latex] (1-6,1) -- (1-6,2.4) node[above=0.5mm] {$\mu$}  ;
\draw[-latex] (1-6,1) -- (2.4-6,1) node[right] {$\nu$};
\coordinate (O) at (-1.8,1) node at (O) [right=-14mm] {$=\hspace{1mm} f_{\lambda} f_{\mu}^{-1} f^{-1}_{\nu}$}  ;
\draw[-latex,dashed] (1,1) -- (2,0) node[below left = -1mm] {$\lambda$};
\draw[-latex,dashed] (1,1) -- (1,2.4) node[above=0.5mm] {$\mu$}  ;
\draw[-latex,dashed] (1,1) -- (-0.4,1) node[left=0.3mm] {$\nu$};
 \end{tikzpicture}
 \label{eq:vertex_anti-vertex_pic}
\end{align}
As mentioned before, the anti-vertex is obtained by flipping the coupling constant, $q \leftrightarrow q^{-1}$.
The Schur function formula \eqref{eq:Schur_transpose} implies that such a flip leads to transposition of the partition $\lambda \leftrightarrow \lambda^\text{T}$, which is similar to the relation between the vertices with clockwise and anticlockwise orientations,
 \be
 \ba
C_{\lambda \mu \nu}(q) = (-1)^{|\lambda|+|\mu|+|\nu|} f_\lambda^{-1} f_\mu^{-1} f_\nu^{-1} C_{\mu^{\rm T} \lambda^{\rm T} \nu^{\rm T}}(q).
 \ea
 \ee
 
The relation \eqref{eq:vertex_anti-vertex} (or graphically Fig. \eqref{eq:vertex_anti-vertex_pic}) leads to the equivalence of the web diagrams with and without the anti-vertex.
For example, we consider the following diagrams:
\begin{align}
  \begin{tikzpicture}[thick,baseline=(current bounding box.center),scale=.5]
   \node at (-3,.5) {($a$)};
   \coordinate (O) at (-.3,0.4) node at (O) {$\o$} ; 
   \coordinate (O) at (2.7,-1) node at (O) {$\nu_1$} ; 
   \coordinate (O) at (2.7,-2) node at (O) {$\nu_2$} ; 
   \coordinate (O) at (2.7,-3) node at (O) {$\nu_3$} ; 
   \coordinate (O) at (-2.3,-4.1) node at (O) {$\o$} ; 
   \coordinate (O) at (.4,-1.4) node at (O) {$Q_1$} ; 
   \coordinate (O) at (-.2,-2.3) node at (O) {$Q_2$} ; 
   \draw (0,0) -- ++(1,-1) -- ++(0,-1) -- ++(-.5,-.5);  
   \draw [dashed] (.5,-2.5) -- ++(-.5,-.5) -- ++(-2,-1);
   \draw (1,-1) -- ++(1,0);
   \draw (1,-2) -- ++(1,0);
   \draw[dashed] (0,-3) -- ++(2,0);   
   \begin{scope}[shift={(10,0)}]
   \coordinate (O) at (-.3,0.4) node at (O) {$\o$} ; 
   \coordinate (O) at (2.7,-1) node at (O) {$\nu_1$} ; 
   \coordinate (O) at (2.7,-2) node at (O) {$\nu_2$} ; 
   \coordinate (O) at (-1.6,-3) node at (O) {$\nu^{\rm T}_3$} ; 
   \coordinate (O) at (-0.,-4.3) node at (O) {$\o$} ; 
   \coordinate (O) at (.4,-1.4) node at (O) {$Q_1$} ; 
   \coordinate (O) at (-.2,-2.3) node at (O) {$Q_2$} ; 
    \node at (-3,.5) {($b$)};
    \draw (0,0) -- ++(1,-1) -- ++(0,-1) -- ++(-1,-1) -- ++(0,-1);
    \draw (1,-1) -- ++(1,0);
    \draw (1,-2) -- ++(1,0);
    \draw (0,-3) -- ++(-1,0);   
   \end{scope}
  \end{tikzpicture}
\end{align}
The left diagram is the chain geometry \eqref{eq:chain_geom}, which is the building block for $\mathrm{U}(2|1)$ gauge theory.
Applying the relation \eqref{eq:vertex_anti-vertex_pic}, we obtain the right diagram which does not contain the anti-vertices.
This is not just a coincidence because the Calabi--Yau geometry corresponding to the right diagram is known to be related to the superalgebra~$\mathfrak{gl}_{2|1}$~\cite{Costello:2018MSRI,Costello:2019COHA,Rapcak:2019wzw}.

More concretely, we can see the equivalence at the level of the building block. It is already given for the configuration $(a)$ in \eqref{LeftPure},
\be\ba
\cZ^{(a)} = \frac{\cZ_{\{ \nu_i \} }^L}{\cZ_{\{ \nu_i =\o\} }^L}\Biggl|_{a=2,b=1,t=q}
\ea\ee
 and, for the configuration $(b)$, it is given by
\be\ba
\cZ^{(b)}_{\{ \nu_i \} }
&= s_{\nu_1^{\rm T}}(q^{-\rho})s_{\nu_2^{\rm T}}(q^{-\rho}) s_{\nu_3}(q^{-\rho}) 
\\
&\qquad\times
\prod_{(i,j)\in \nu_1} \f{(1-Q_1 Q_2 q^{i+j-\nu_{1,i}-\nu_{3,j}^{\rm T}-1})}{(1-Q_1 q^{i+j-\nu_{1,i}-\nu_{2,j}^{\rm T}-1})}
\prod_{(i,j)\in \nu_2} \f{(1-Q_2 q^{i+j-\nu_{2,i}-\nu_{3,j}^{\rm T}-1})}{(1-Q_1 q^{-i-j+\nu_{2,i}+\nu_{1,j}^{\rm T}+1})}
\\
&\qquad\times
\prod_{(i,j)\in \nu_3} (1-Q_1 Q_2  q^{-i-j+\nu_{3,i}+\nu_{1,j}^{\rm T}+1})(1-Q_2 q^{-i-j+\nu_{3,i}+\nu_{2,j}^{\rm T}+1}).
\ea\ee
By using the analytic continuation formula \eqref{analyticunref}, we find the agreement between these expressions.


\section{Summary and Discussion}\label{Disc}

In this paper, we have proposed a new topological vertex formalism including the anti-vertex, which is motivated by the supergroup gauge theory.
We have computed several topological string amplitudes, and shown their agreement with the supergroup gauge theory partition function.
We have then pointed out the one-to-many correspondence between the gauge theory and the Calabi--Yau geometry, which is a specific property of the supergroup theory.
We have also discussed the relation between the ordinary vertex and the anti-vertex through the analytic continuation.
This is consistent with the known argument on the supergroup theory.

Although we have introduced the anti-vertex and the associated web diagrams, its geometric interpretation is not yet obvious.
In order to engineer the $\mathrm{SU}(n)$ gauge symmetry, the $A_{n-1}$ singularity involved in the Calabi--Yau geometry is utilized to reproduce the gauge theory result.
This implies that we should consider an exotic singularity, i.e., $A_{m-1|n-1}$ singularity, to realize the $\mathrm{SU}(m|n)$ gauge theory.
Actually the Taub-NUT geometry with the negative charge discussed in~\cite{DHJV} would be interpreted as an example with such a super-type singularity, which also suggests a super analog of the McKay--Nakajima correspondence~\cite{McKay:1981,Nakajima:1994qu,Nakajima:1998DM}.
Another issue in the anti-vertex formalism is the framing factor.
In this paper, we have mainly considered the situation in which the framing factor does not play a crucial role.
In order to apply our formalism to more generic Calabi--Yau geometries, it would be important to provide a precise prescription to fix the framing factor.

We also remark a possible application of the formalism presented in this paper to the quiver gauge theory having the fermionic node, which should be interpreted as the base/fibre dual; S-dual; spectral dual to the situation studied in this paper.
It would be interesting to apply the topological vertex/anti-vertex formalism to explore such a new configuration in gauge theory~\cite{Orlando:2010uu,Nekrasov:2018gne,Zenkevich:2018fzl}.

In addition to the partition function itself, it is also interesting to study the situation in the presence of the defect operators.
For example, the $qq$-character~\cite{Bourgine:2015szm,Nekrasov:2015wsu,Kimura:2015rgi} introduced in gauge theory has a realization using a codimension-4 defect, whose topological string setup has been proposed in~\cite{Kimura:2017auj}.
Another important class of the defect, which is a codimension-2 surface defect, can be also discussed in the context of topological string~\cite{Dimofte:2010tz, Taki:2010bj, Mori:2016qof} through the geometric transition~\cite{Gopakumar:1998ki}.
It would be a natural generalization to apply this formalism to the present case involving the anti-vertex, and its justification from the gauge theory point of view would be also an interesting issue.

\subsubsection*{Acknowledgements}

The work of TK was supported in part by the French ``Investissements d'Avenir'' program, project ISITE-BFC (No.~ANR-15-IDEX-0003), JSPS Grant-in-Aid for Scientific Research (No.~JP17K18090), the MEXT-Supported Program for the Strategic Research Foundation at Private Universities ``Topological Science'' (No.~S1511006), JSPS Grant-in-Aid for Scientific Research on Innovative Areas ``Topological Materials Science'' (No.~JP15H05855), and ``Discrete Geometric Analysis for Materials Design'' (No.~JP17H06462).
The work of YS was support by a grant from the NSF of China with Grant No:~11947301.


\appendix
\section{Notation, Definition, and Formulae}\label{ND}

\subsection{Notation and Definition}
Here we summarize the definitions and notations used in this paper.
For a given partition $\lambda$, we denote the transposed partition by $\lambda^\text{T}$.
The refined topological vertex is given by
\be\ba
&C_{\lambda \mu \nu}(t,q) = t^{-\frac{||\mu^{t}||^{2}}{2}}q^{\frac{||\mu||^2 + ||\nu||^{2}}{2}} \tilde{Z}_{\nu}(t,q)
 \sum_{\eta}\Bigl(\frac{q}{t}\Bigr)^{\frac{|\eta| + |\lambda| - |\mu|}{2}}  s_{\lambda^\text{T}/\eta}(t^{-\rho}q^{-\nu})s_{\mu/\eta}(t^{-\nu^\text{T}}q^{-\rho}), 
 \\
 &\tilde{Z}_{\nu}(t,q) = \prod_{(i,j) \in \nu}(1-q^{\nu_{i}-j}t^{\nu_{j}^\text{T} -i +1})^{-1},
 \ea\ee
 where
\be\ba
&|\mu|=\sum_{i=1}^{l(\mu)}\mu_{i} , \qquad
||\mu||^{2}=\sum_{i=1}^{l(\mu)}\mu^{2}_{i}, \qquad 
\rho:= \left\{ -i+\frac{1}{2} \right\}_{i = 1,\ldots,\infty} ,
\ea\ee
and $s_{\mu/\eta}(x)$ is the so-called skew-Schur function. The third component of the refined topological vertex is called as the preferred direction.
The refined topological vertex is simplified in the unrefined limit $t = q$,
\be\ba
C_{\lambda \mu \nu}(q,q) = q^{\frac{\kappa_{\mu}}{2}}s_{\nu^{\rm T}}(q^{-\rho})
\sum_{\eta} s_{\lambda^\text{T}/\eta}(q^{-\nu-\rho})s_{\mu/\eta}(q^{-\rho-\nu^\text{T}}).
\ea\ee
To glue two vertices, we need to insert the framing factors,
\be
\ba
&f_\mu (t,q) = (-1)^{|\mu|} t^{\f{||\mu^{\rm T}||^2}{2}} q^{-\f{||\mu||^2}{2}} \qquad \text{for the preferred direction,}
\\
&\tilde{f}_\mu (t,q) = (-1)^{|\mu|} \left(\f{t}{q}\right)^{\f{|\mu|}{2}} t^{\f{||\mu^{\rm T}||^2}{2}} q^{-\f{||\mu||^2}{2}} \qquad \text{for other directions,}
\ea
\ee
which are the same under the unrefined limit.


\subsection{Schur function formulae}
The (refined) topological vertex is given using the Schur function. Here we summarize some formulae for the Schur function that we use for the caluculation:
\begin{subequations}
\begin{align}
s_{\lambda/\mu}(\alpha \bold{x})
&= \alpha^{|\lambda|-|\mu|}s_{\lambda/\mu}(\bold{x}),
\\
s_{\alpha} (q^{\rho+\beta}) &= (-1)^{|\alpha|}s_{\alpha^{\rm T}}(q^{-\rho-\beta^{\rm T}}), \label{eq:Schur_transpose}
\\
\sum_{\eta}s_{\eta/\lambda}(\bold{x})s_{\eta/\mu}(\bold{y})
&= \prod_{i,j=1}^{\infty}(1-x_{i}y_{j})^{-1}
\sum_{\eta}s_{\mu/\eta}(\bold{x})s_{\lambda/\eta}(\bold{y}),
\\
\sum_{\eta}s_{\eta^\text{T}/\lambda}(\bold{x})s_{\eta/\mu}(\bold{y})
&= \prod_{i,j=1}^{\infty}(1+x_{i}y_{j})
\sum_{\eta}s_{\mu^\text{T}/\eta^\text{T}}(\bold{x})s_{\lambda^\text{T}/\eta}(\bold{y}).
\end{align}
\end{subequations}

To relate the topological string partition function to the Nekrasov partition function, we sometimes use following normalization formulae:
\begin{subequations}
\begin{align}
\prod_{i,j=1}^{\infty} \frac{1-Qq^{\nu_{i}-j}t^{\mu_{j}^\text{T}-i+1}}{1-Qq^{-j}t^{-i+1}}
 &= \prod_{(i,j) \in \nu}(1-Qq^{\nu_{i}-j}t^{\mu_{j}^\text{T}-i+1})\prod_{(i,j) \in \mu}(1-Qq^{-\mu_{i}+j-1}t^{-\nu_{j}^\text{T}+i}),
\\
\prod_{i,j=1}\f{1-Q t^{i+\mu_j^{\rm T}} q^{-j-\nu_i}}{1-Q t^i q^{-j}}
&=
\prod_{j=1}^{\mu_1} \prod_{i=1}^{\nu_1^{\rm T}} \f{1-Q t^{i+\mu_j^{\rm T}} q^{-j-\nu_i}}{1-Q t^i q^{-j}}
 \prod_{(i,j)\in \mu} \f{1}{1-Q t^{\nu^{\rm T}_1+i} q^{-j}} \prod_{(i,j) \in \nu} \f{1}{1 - Q t^{i} q^{-j-\mu_1}}.
\end{align}
\end{subequations}
The first formula is frequently used to compare the topological string amplitude with the gauge theory partition function, whereas the second one is particularly used for the anti-vertex formalism.

\subsection{Analytic continuation formula}
To show the equivalence of the partition functions of supergroup gauge theories and usual gauge theories, we use following formula,
\be\ba
\prod_{i,j=1}^\infty (1-Q q^{i+j-\beta_i -\gamma_j^{\rm T}-1}) = \prod_{i,j=1}^\infty \f{1}{1-Q q^{-i+j+\beta_i^{\rm T}-\gamma_j^{\rm T}}}.
\label{analyticunref}
\ea\ee
To show the agreement of the partition functions for more than two web diagrams giving U(2$|$1) gauge theory, we use following formula,
\be\ba
\prod_{i,j=1}^\infty (1-Q t^{i-\gamma_j^{\rm T}-\f{1}{2}} q^{j-\beta_i-\f{1}{2}} )= \prod_{i,j=1}^\infty (1-Q t^{-i+\beta_j^{\rm T}+\f{1}{2}} q^{-j+\gamma_i+\f{1}{2}} ).
\label{analyticref}
\ea\ee



\providecommand{\href}[2]{#2}\begingroup\raggedright\endgroup


\end{document}